\documentclass{aastex62}

\received{}
\revised{}
\accepted{}
\submitjournal{ApJ}

\shorttitle{Chemical composition in three MYSOs associated with 6.7 GHz methanol masers}
\shortauthors{Taniguchi et al.}

\begin{document}

\title{Chemical Diversity in Three Massive Young Stellar Objects associated with 6.7 GHz CH$_{3}$OH Masers}

\correspondingauthor{Kotomi Taniguchi}
\email{kotomi.taniguchi@nao.ac.jp}

\author{Kotomi Taniguchi}
\altaffiliation{Research Fellow of Japan Society for the Promotion of Science}
\altaffiliation{Present address: National Astronomical Observatory of Japan, Osawa, Mitaka, Tokyo 181-8588, Japan}
\affiliation{Department of Astronomical Science, School of Physical Science, SOKENDAI (The Graduate University for Advanced Studies), Osawa, Mitaka, Tokyo 181-8588, Japan}
\affiliation{Nobeyama Radio Observatory, National Astronomical Observatory of Japan, Minamimaki, Minamisaku, Nagano 384-1305, Japan}

\author{Masao Saito}
\affiliation{National Astronomical Observatory of Japan, Osawa, Mitaka, Tokyo 181-8588, Japan}
\affiliation{Department of Astronomical Science, School of Physical Science, SOKENDAI (The Graduate University for Advanced Studies), Osawa, Mitaka, Tokyo 181-8588, Japan}

\author{Liton Majumdar}
\affiliation{Jet Propulsion Laboratory, California Institute of Technology, 4800 Oak Grove Drive, Pasadena, CA 91109, USA}

\author{Tomomi Shimoikura}
\affiliation{Department of Astronomy and Earth Sciences, Tokyo Gakugei University, Nukuikitamachi, Koganei, Tokyo 184-8501, Japan}

\author{Kazuhito Dobashi}
\affiliation{Department of Astronomy and Earth Sciences, Tokyo Gakugei University, Nukuikitamachi, Koganei, Tokyo 184-8501, Japan}

\author{Hiroyuki Ozeki}
\affiliation{Department of Environmental Science, Faculty of Science, Toho University, Miyama, Funabashi, Chiba 274-8510, Japan}

\author{Fumitaka Nakamura}
\affiliation{National Astronomical Observatory of Japan, Osawa, Mitaka, Tokyo 181-8588, Japan}
\affiliation{Department of Astronomical Science, School of Physical Science, SOKENDAI (The Graduate University for Advanced Studies), Osawa, Mitaka, Tokyo 181-8588, Japan}

\author{Tomoya Hirota}
\affiliation{National Astronomical Observatory of Japan, Osawa, Mitaka, Tokyo 181-8588, Japan}
\affiliation{Department of Astronomical Science, School of Physical Science, SOKENDAI (The Graduate University for Advanced Studies), Osawa, Mitaka, Tokyo 181-8588, Japan}

\author{Tetsuhiro Minamidani}
\affiliation{Nobeyama Radio Observatory, National Astronomical Observatory of Japan, Minamimaki, Minamisaku, Nagano 384-1305, Japan}
\affiliation{Department of Astronomical Science, School of Physical Science, SOKENDAI (The Graduate University for Advanced Studies), Osawa, Mitaka, Tokyo 181-8588, Japan}

\author{Yusuke Miyamoto}
\altaffiliation{Present address: National Astronomical Observatory of Japan, Osawa, Mitaka, Tokyo 181-8588, Japan}
\affiliation{Nobeyama Radio Observatory, National Astronomical Observatory of Japan, Minamimaki, Minamisaku, Nagano 384-1305, Japan}

\author{Hiroyuki Kaneko}
\affiliation{Nobeyama Radio Observatory, National Astronomical Observatory of Japan, Minamimaki, Minamisaku, Nagano 384-1305, Japan}



\begin{abstract}
We have carried out observations in the 42$-$46 and 82$-$103 GHz bands with the Nobeyama 45-m radio telescope, and in the 338.2$-$339.2 and 348.45$-$349.45 GHz bands with the ASTE 10-m telescope toward three high-mass star-forming regions containing massive young stellar objects (MYSOs), G12.89+0.49, G16.86$-$2.16, and G28.28$-$0.36.
We have detected HC$_{3}$N including its $^{13}$C and D isotopologues, CH$_{3}$OH, CH$_{3}$CCH, and several complex organic molecules (COMs).
Combining our previous results of HC$_{5}$N in these sources, we compare the $N$(HC$_{5}$N)/$N$(CH$_{3}$OH) ratios in the three observed sources.
The ratio in G28.28$-$0.36 is derived to be $0.091^{+0.109}_{-0.039}$, which is higher than that in G12.89+0.49 by one order of magnitude, and that in G16.86$-$2.16 by a factor of $\sim 5$.
We investigate the relationship between the $N$(HC$_{5}$N)/$N$(CH$_{3}$OH) ratio and the $N$(CH$_{3}$CCH)/$N$(CH$_{3}$OH) ratio.
The relationships of the two column density ratios in G28.28$-$0.36 and G16.86$-$2.16 are similar to each other, while HC$_{5}$N is less abundant when compared to CH$_{3}$CCH in G12.89+0.49.   
These results imply a chemical diversity in the lukewarm ($T \sim 20-30$ K) envelope around MYSOs.
Besides, several spectral lines from complex organic molecules, including very-high-excitation energy lines, have been detected toward G12.89+0.49, while the line density is significantly low in G28.28$-$0.36.
These results suggest that organic-poor MYSOs are surrounded by a carbon-chain-rich lukewarm envelope (G28.28$-$0.36), while organic-rich MYSOs, namely hot cores, are surrounded by a CH$_{3}$OH-rich lukewarm envelope (G12.89+0.49 and G16.86$-$2.16).
\end{abstract}




\section{Introduction} \label{sec:intro}

Molecules are a unique and powerful tool in astronomy.
They provide an excellent diagnosis of the physical conditions and processes in the regions where they reside.
Progress in this field called astrochemistry is mainly driven by observations from various single-dish and interferometers at millimeter/sub-millimeter wavelengths along with space telescopes at mid- and far-infrared wavelengths \citep[e.g.,][]{2017arXiv171005940V}.

Unsaturated carbon-chain molecules tend to be abundant in young low-mass starless cores and deficient in low-mass star-forming cores \citep{1992apj...392...551, 2009apj...699...585}, because they are mainly formed from ionized carbon (C$^{+}$) and atomic carbon (C) via ion-molecule reactions in the early stage of molecular clouds and destroyed mainly by reaction with oxygen atoms and depleted onto dust grains in the late stage.
Carbon-chain molecules have been found to be abundant around two low-mass protostars; IRAS 04368+2557 in L1527 \citep{2008ApJ...672..371S} and IRAS 15398$-$3359 in Lupus \citep{2009ApJ...697..769S}.
In these protostars, carbon-chain molecules are newly formed from CH$_{4}$ evaporated from dust grains in the lukewarm ($T \sim 20-30$ K) gas \citep{2008ApJ...672..371S}.
Such a carbon-chain chemistry around the low-mass protostars was named warm carbon chain chemistry \citep[WCCC,][]{2008ApJ...672..371S}.
The difference between hot corino chemistry and WCCC is considered to be brought by the different timescale of the starless core phase; the long and short starless core phases lead to hot corino and WCCC sources, respectively \citep{2008ApJ...672..371S}.
Recently, sources possessing both hot corino and WCCC characteristics have been found and high-spatial-resolution observations showed that the spatial distributions of carbon-chain molecules and COMs are different with each other \citep[e.g.,][]{2016ApJ...830L..37I}.

Saturated complex organic molecules (COMs), organic species consisting of more than six atoms and being rich in hydrogen \citep{2009ARA&A..47..427H}, are classically known to be abundant in the dense and hot ($n > 10^{6}$ cm$^{-3}$, $T \geq 100$ K) gas around young stellar objects (YSOs).
Besides, COMs also have been found in the gas phase before ice thermally evaporates at temperatures above 100 K \citep{2014ApJ...795L...2V}.
At these low temperatures, COMs can be desorbed from icy grain mantles via different types of non-thermal desorption process such as: (1) the cosmic-ray desorption mechanism \citep{2014MNRAS.440.3557R}, (2) the chemical desorption mechanism \citep[desorption due to the exothermicity of surface reactions;][]{2007A&A...467.1103G}, and (3) the photo-desorption \citep{2016MNRAS.459.3756R}. 
Role of barrier-less gas phase reactions to form COMs was also proposed recently by \citet{2015MNRAS.449L..16B}.
Recent observations show the presence of some COMs, methylformate (HCOOCH$_{3}$), dimethyl ether (CH$_{3}$OCH$_{3}$), and methyl cyanide (CH$_{3}$CN), and the complex radical methoxy (CH$_{3}$O) in regions where the dust temperature is less than 30 K; pre-stellar cores \citep{2012A&A...541L..12B, 2014ApJ...795L...2V, 2016A&A...594A.117P} and cold envelopes of low-mass protostars \citep{2010ApJ...716..825O, 2012ApJ...759L..43C, 2014ApJ...791...29J}.
Grain-surface chemistry certainly plays a role, for example in forming hydrogenated species during the prestellar phase \citep{1982A&A...114..245T, 2012A&ARv..20...56C}, but not necessarily in the formation of all COMs.

Not only in the low-mass star-forming regions, new questions arise in the chemistry around massive young stellar objects (MYSOs).
\citet{2015A&A...576A..45F} compared the chemistry between organic-poor MYSOs and organic-rich MYSOs, namely hot cores.
They suggested that hot cores are not required to form COMs and temperature and initial ice composition possibly affect complex organic distributions around MYSOs.
\citet{2014MNRAS...443...2252} detected HC$_{5}$N, the second shortest cyanopolyynes (HC$_{2n+1}$N, $n=1,2,3,...$), in 35 hot cores associated with the 6.7 GHz methanol masers, which give us the exact positions of MYSOs \citep{2013MNRAS.431.1752U}.
However, there remained the possibility that the emission of HC$_{5}$N comes from the outer cold molecular clouds or other molecular clouds in the large single-dish beam ($\sim 0.95$\arcmin).
\citet{2017ApJ...844...68T} carried out observations toward four MYSOs where \citet{2014MNRAS...443...2252} detected HC$_{5}$N, using the Green Bank 100-m telescope (GBT) and the Nobeyama 45-m radio telescope.
The four target sources were selected adding three criteria mentioned in Section \ref{sec:obs} and we chose three sources showing the highest HC$_{5}$N peak intensities (G12.89+0.49, G16.86$-$2.16, and G28.28$-$0.36) and a source showing the low HC$_{5}$N peak intensity with the high CH$_{3}$CN peak intensities (G10.30$-$0.15).
They detected the high-excitation-energy ($E_{\rm {u}}/k \sim 100$ K) lines of HC$_{5}$N, which cannot be detected if HC$_{5}$N exists in the cold dark clouds, in the three sources (G12.89+0.49, G16.86$-$2.16, and G28.28$-$0.36) and confirmed that HC$_{5}$N exists in the warm gas around MYSOs.
Therefore, carbon-chain molecules seem to be formed in the lukewarm gas around MYSOs, as well as WCCC sources in the low-mass star-forming regions.
At the present stage, we do not know the relationships between carbon-chain molecules and COMs around MYSOs.

In this paper, we report the observational results in the 42$-$46, 82$-$103, 338.2$-$339.2 and 348.45$-$349.45 GHz bands obtained with the Nobeyama 45-m radio telescope and the Atacama Submillimeter Telescope Experiment (ASTE) 10-m telescope toward three MYSOs, G12.89+0.49, G16.86$-$2.16, and G28.28$-$0.36.
We derive the rotational temperatures and beam-averaged column densities of HC$_{3}$N, CH$_{3}$OH, and CH$_{3}$CCH (Section \ref{sec:ana}).  
We compare the spectra and the chemical composition among the three sources, combining with previous HC$_{5}$N data \citep{2017ApJ...844...68T}, in order to investigate the relationship between carbon-chain species and COMs in high-mass star-forming regions (Section \ref{sec:discuss}).

\section{Observations} \label{sec:obs}

The observations presented in this paper were conducted in the Nobeyama 45-m radio telescope and the ASTE joint observation program (Proposal ID: JO161001, PI: Kotomi Taniguchi, 2016-2017 season).
The observing parameters of each frequency band are summarized in Table \ref{tab:obs}.
The Nobeyama 45-m telescope data have been partly published in \citet{2017ApJ...844...68T}.

The source selection criteria were described in \citet{2017ApJ...844...68T} as follows:
\begin{enumerate}
\item The source declination is above $-21$\arcdeg,
\item the distance ($D$) is within 3 kpc, and
\item CH$_{3}$CN was detected \citep[$\int T_{\rm {mb}}dv > 0.5$ K km s$^{-1}$ for $J_{K}=5_{0}-4_{0}$ line,][]{2006MNRAS...367...553}.
\end{enumerate}
Eight sources in the HC$_{5}$N-detected source list of \citet{2014MNRAS...443...2252} meet the above three criteria. 
We chose three sources among the eight selected sources which show the highest peak intensities of HC$_{5}$N ($T_{\rm {mb}} > 120$ mK).
Table \ref{tab:source} summarizes the properties of our three target sources.
The observed positions correspond to the 6.7 GHz methanol maser positions, which show exact positions of the MYSOs \citep{2013MNRAS.431.1752U}.

\floattable
\begin{deluxetable}{ccccccc}
\tablecaption{Observing parameters\label{tab:obs}}
\tablewidth{0pt}
\tablehead{
\colhead{Frequency} & \colhead{Telescope} & \colhead{Beam size} & \colhead{$\eta_{\rm {mb}}$} & \colhead{$T_{\rm {sys}}$} & \colhead{$\Delta \nu$} & \colhead{$T_{\rm {rms}}$\tablenotemark{a}} \\
\colhead{(GHz)} & \colhead{} & \colhead{($^{\prime\prime}$)} & \colhead{(\%)} & \colhead{(K)} & \colhead{(kHz)} & \colhead{(mK)}
}
\startdata
42$-$46 & Nobeyama & 37 & 71 & 120$-$150 & 122.07 & 6$-$14 \\
82$-$103 & Nobeyama & 18\tablenotemark{b} & 54\tablenotemark{b} & 120$-$200 & 244.14 & 3$-$6 \\
338$-$349.4 & ASTE & 22 & 60 & 300$-$700 & 1000 &16$-$24  \\
\enddata
\tablenotetext{a}{In $T_{\rm {A}}^{*}$ scale.}
\tablenotetext{b}{The values are at the 86 GHz.}
\end{deluxetable}
\floattable
\begin{deluxetable}{ccccccc}
\tablecaption{Properties of our target sources \label{tab:source}}
\tablewidth{0pt}
\tablehead{
\colhead{Source} & \colhead{R.A.\tablenotemark{a}} & \colhead{Decl.\tablenotemark{a}} &\colhead{$D$} & \colhead{$V_{\rm {LSR}}$\tablenotemark{a}} & \multicolumn{2}{c}{Other Association\tablenotemark{b}} \\
\cline{6-7}
\colhead{} & \colhead{(J2000)} & \colhead{(J2000)} & \colhead{(kpc)} & \colhead{(km s$^{-1}$)} & \colhead{UC\ion{H}{2}\tablenotemark{a}} & \colhead{outflow} 
}
\startdata
G12.89+0.49 & 18$^{\rm h}$11$^{\rm m}$51\fs4 & -17\arcdeg31\arcmin30\arcsec & 2.50\tablenotemark{c} & 33.3 & N & Y\tablenotemark{d} \\
G16.86$-$2.16 & 18$^{\rm h}$29$^{\rm m}$24\fs4 & -15\arcdeg16\arcmin04\arcsec & 1.67\tablenotemark{d} & 17.8 & N & Y\tablenotemark{d} \\
G28.28$-$0.36 & 18$^{\rm h}$44$^{\rm m}$13\fs3 & -04\arcdeg18\arcmin03\arcsec & 3.0\tablenotemark{e} & 48.9 & Y & Y\tablenotemark{f} \\
\enddata
\tablenotetext{a}{\citet{2006MNRAS...367...553}}
\tablenotetext{b}{The symbols of ``Y" and ``N" represent detection and non-detection, respectively. 
``UC\ion{H}{2}" indicates an ultracompact \ion{H}{2} region lying within a radius of $19\arcsec$. 
The 6.7 GHz methanol masers are associated with all of the four sources.}
\tablenotetext{c}{\citet{2014apj...783...130}}
\tablenotetext{d}{\citet{2016aj...152...92L}}
\tablenotetext{e}{\citet{2014MNRAS...443...2252}}
\tablenotetext{f}{\citet{2008AJ...136...2391}}
\end{deluxetable}

\subsection{Observations with the Nobeyama 45-m Radio Telescope} \label{sec:obsNRO}

We carried out observations with the Nobeyama 45-m radio telescope from 2017 January to March.
We employed the position-switching mode.
The integration time was 20 seconds per on-source and off-source positions.
The on-source positions are summarized in Table \ref{tab:source} and the off-source positions were set to be $+15\arcmin$ away in declination. 
The total integration time is $\sim 1$ hr and 2$-$4.5 hr in the 42$-$46 GHz and 82$-$103 GHz band observations, respectively.

The Z45 receiver \citep{2015PASJ...67..117N} and the TZ receiver \citep{2013PASP..125..252N} were used in the observations at 42$-$46 GHz and 82$-$103 GHz, respectively.
The main beam efficiency ($\eta_{\rm {mb}}$) and the beam size (HPBW) at 43GHz were 71\% and $37\arcsec$, respectively.
The main beam efficiency and the beam size at 86 GHz were 54\% and $18\arcsec$, respectively.
The system temperatures were 120$-$150 K and 120$-$200 K during the observations at 42$-$46 GHz and 82$-$103 GHz, respectively.
We used the SAM45 FX-type digital correlator \citep{2012PASJ...64...29K} in frequency settings whose bandwidths and resolutions are 500 MHz and 122.07 kHz for the Z45 observations, and 1000 MHz and 244.14 kHz for the TZ observations, respectively.
The frequency resolutions correspond to the velocity resolution of $\sim 0.85$ km s$^{-1}$.

The telescope pointing was checked every 1.5 hr by observing the SiO maser line ($J=1-0$; 43.12203 GHz) from OH39.7+1.5 at ($\alpha_{2000}$, $\delta_{2000}$) = (18$^{\rm h}$56$^{\rm m}$03\fs88, +06\arcdeg38\arcmin49\farcs8).
We used the Z45 receiver for the pointing check during the 42$-$46 GHz band observations and the H40 receiver for the pointing check during the 82$-$103 GHz band observations.
The pointing error was less than $3\arcsec$.

The rms noises are 6$-$14 and 3$-$6 mK in $T_{\rm {A}}^{*}$ scale in the 42$-$46 GHz and 82$-$103 GHz bands, respectively.
The baseline was fitted with a linear function.
The absolute flux calibration error is approximately 10\%.

\subsection{Observations with the ASTE 10-m Telescope} \label{sec:obsASTE}

The observations with the ASTE 10-m telescope were conducted in 2016 September and October.
The DASH345 receiver and the WHSF FX-type digital spectrometer \citep{2008PASJ...60..857I, 2008PASJ...60..315O} were used.
The observed frequency ranges are 338.2$-$339.2 and 348.45$-$349.45 GHz.
The main beam efficiency and beam size were 60\% and $22\arcsec$, respectively.
The system temperatures were between 300 and 700 K, depending on the elevation and weather conditions.
The frequency setting was 2048 MHz bandwidth and 1 MHz frequency resolution.
The frequency resolution corresponds to 0.86 km s$^{-1}$, which is almost equal to that of observations with the Nobeyama 45-m telescope.
The total integration time is approximately 1-- 2.5 hr, which is different among sources.

We checked the telescope pointing every 2 hr by observing the $^{12}$CO ($J = 3-2$) line from W Aql at ($\alpha_{2000}$, $\delta_{2000}$) = (19$^{\rm h}$15$^{\rm m}$23\fs35, -07\arcdeg02\arcmin50\farcs3).
The pointing error was less than $2\arcsec$.

The rms noise levels in the line-free regions are 23$-$24, 19$-$20, and 16$-$17 mK in $T_{\rm {A}}^{*}$ scale in G12.89+0.49, G16.86$-$2.16, and G28.28$-$0.36, respectively.
Some scans were excluded due to bad baselines. 
A linear fit was applied for baseline subtraction.
The absolute flux calibration error is approximately 10\%.

\section{Results}

We conducted data reduction using the Java Newstar software, an astronomical data analyzing system of the Nobeyama 45-m telescope and the ASTE 10-m telescope.

\subsection{Observational Results with the Nobeyama 45-m Radio Telescope} \label{sec:resNRO}

\begin{figure}
\figurenum{1}
\plotone{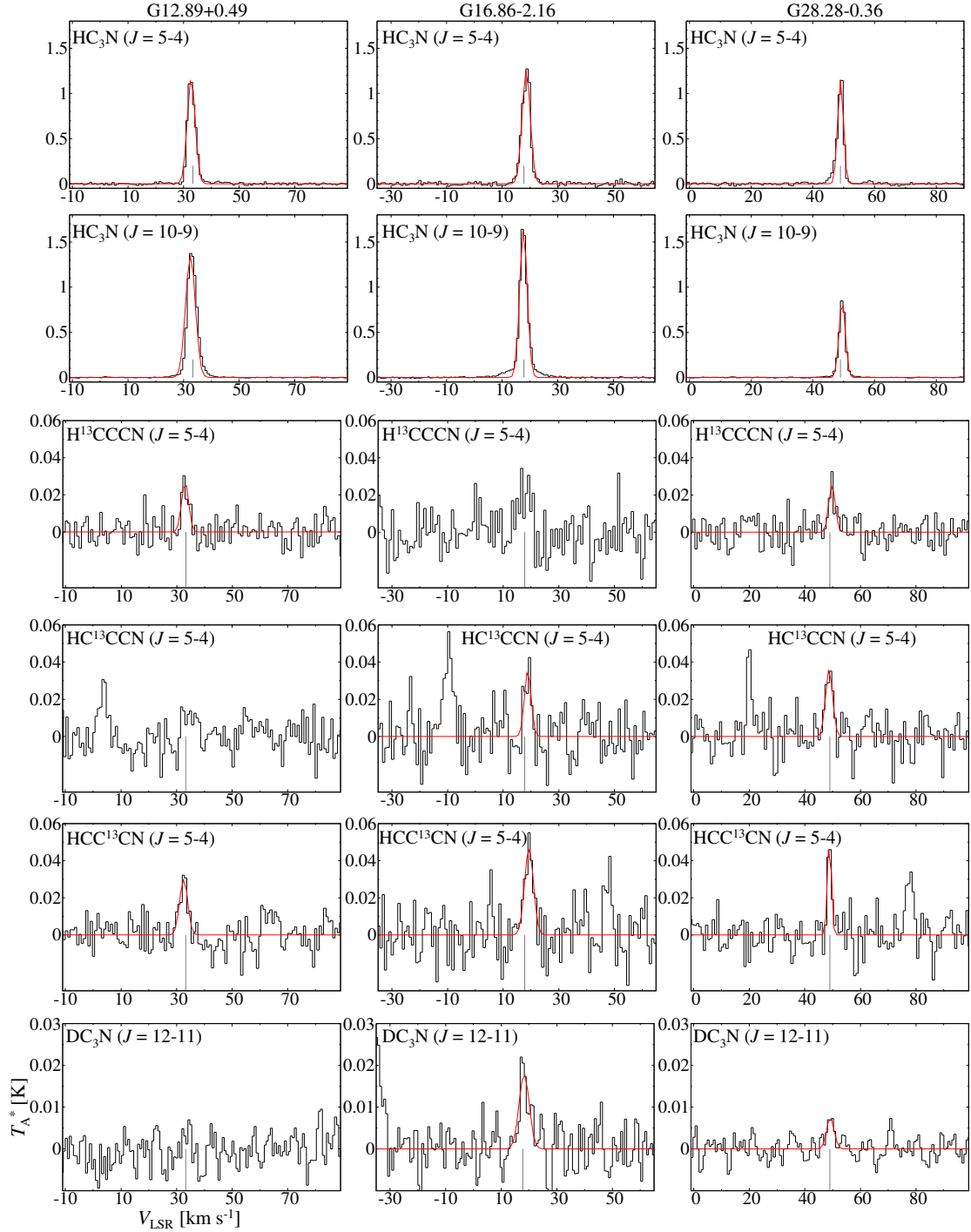}
\caption{Spectra of the main isotopologue HC$_{3}$N and its $^{13}$C and D isotopologues in the three sources. The red lines show the Gaussian fitting results and gray vertical lines show the systemic velocity for each source.\label{fig:f1}}
\end{figure}

Figure \ref{fig:f1} shows the spectra of the main isotopologue HC$_{3}$N and its $^{13}$C and D isotopologues in the three sources obtained with the Nobeyama 45-m radio telescope.
We fitted the spectra with a Gaussian function and obtained the spectral line parameters as summarized in Table \ref{tab:resHC3N}.
Two rotational transitions, $J=5-4$ and $10-9$, of the main isotopologue are in the observed frequency bands, and they were detected from all of the three sources.
Its three $^{13}$C isotopologues have been detected from G28.28$-$0.36 with a signal-to-noise (S/N) ratio above 4.
HC$^{13}$CCN was not detected in G12.89+0.49, whereas H$^{13}$CCCN was not detected in G16.86$-$2.16.
DC$_{3}$N was detected in G28.28$-$0.36 with a S/N ratio of 3 and in G16.86$-$2.16 with a S/N ratio above 4.
The $V_{\rm {LSR}}$ values agree with the systemic velocities of each source (Table \ref{tab:source}).

\begin{figure}
\figurenum{3}
\plotone{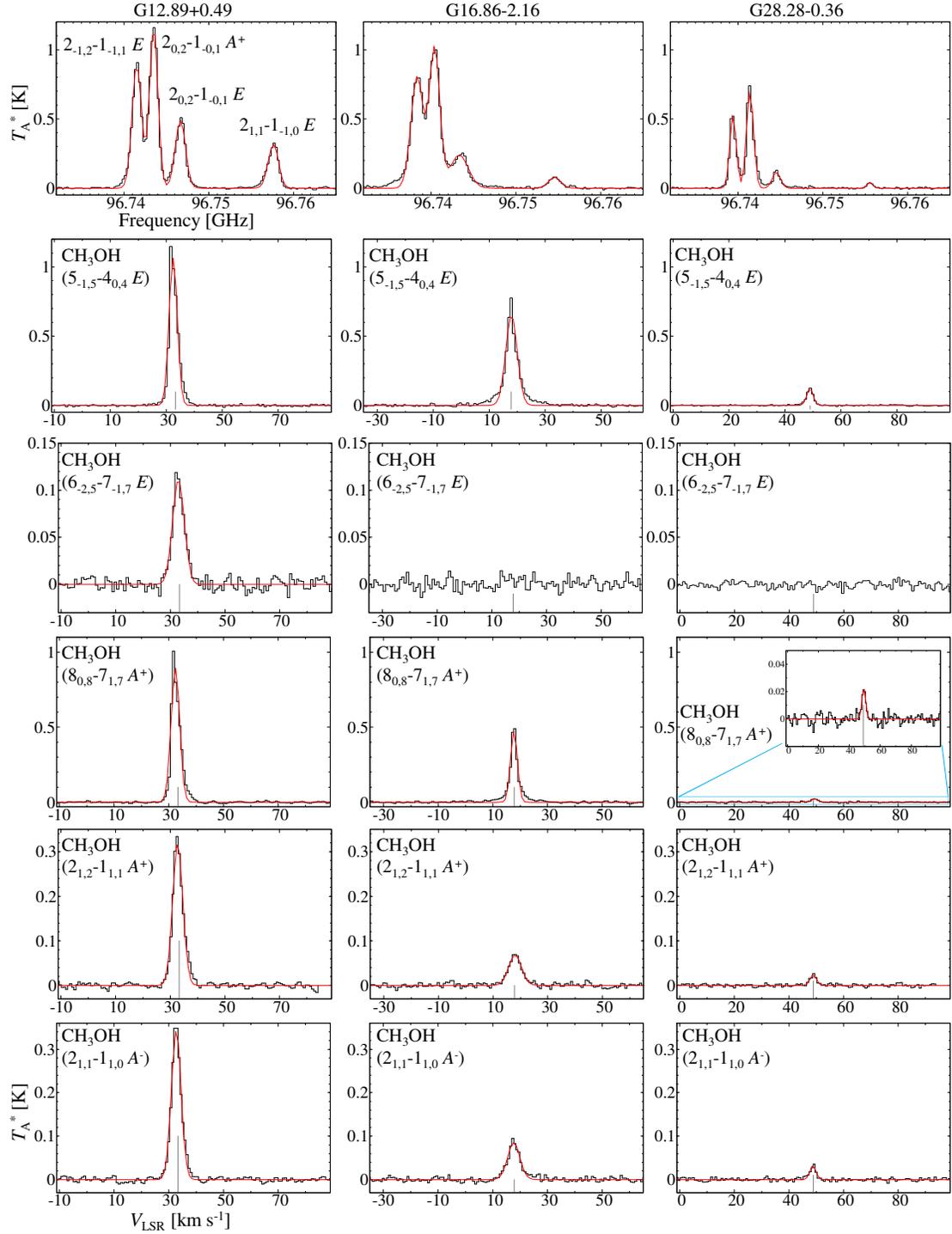}
\caption{Spectra of CH$_{3}$OH in the three sources. The red lines show the Gaussian fitting results and gray vertical lines show the systemic velocity for each source.\label{fig:f3}}
\end{figure}

The line profiles of the main isotopologue show wing emission, suggesting that HC$_{3}$N also exists in the molecular outflow \citep[e.g.,][]{2015ApJS..221...31S, 2018ApJ...854..133T}.
Such wing emission is most prominent in G16.86$-$2.16, where the blue and red components are clearly detected.
The red and blue components are prominent in G12.89+0.49 and G28.28$-$0.36, respectively.
These features of wing emission of HC$_{3}$N are similar to those of CH$_{3}$OH (Figure \ref{fig:f3}), as mentioned later.
Hence, the origin of molecular outflows is plausible.

\floattable
\rotate
\begin{deluxetable}{lllcccccccccccccc}
\tabletypesize{\scriptsize}
\tablecaption{Spectral line parameters of the main isotopologue HC$_{3}$N and its $^{13}$C and D isotopologues in the three sources with the Nobeyama 45-m telescope \label{tab:resHC3N}}
\tablewidth{0pt}
\tablehead{
\colhead{} & \colhead{} & \colhead{}  & \multicolumn{4}{c}{G12.89+0.49} & \colhead{} & \multicolumn{4}{c}{G16.86$-$2.16} & \colhead{} & \multicolumn{4}{c}{G28.28$-$0.36} \\
\cline{4-7}\cline{9-12}\cline{14-17}
\colhead{Species} & \colhead{Frequency\tablenotemark{a}} & \colhead{$E_{\rm {u}}/k$} & \colhead{$T_{\rm {mb}}$} & \colhead{FWHM} & \colhead{$\int T_{\mathrm {mb}}dv$} & \colhead{$V_{\rm{LSR}}$\tablenotemark{b}} & \colhead{} & \colhead{$T_{\rm {mb}}$} & \colhead{FWHM} & \colhead{$\int T_{\mathrm {mb}}dv$} & \colhead{$V_{\rm{LSR}}$\tablenotemark{b}} & \colhead{} & \colhead{$T_{\rm {mb}}$} & \colhead{FWHM} & \colhead{$\int T_{\mathrm {mb}}dv$} & \colhead{$V_{\rm{LSR}}$\tablenotemark{b}} \\
\colhead{Transition} & \colhead{(GHz)} & \colhead{(K)} & \colhead{(K)} & \colhead{(km s$^{-1}$)} & \colhead{(K km s$^{-1}$)} & \colhead{(km s$^{-1}$)} & \colhead{} & \colhead{(K)} & \colhead{(km s$^{-1}$)} & \colhead{(K km s$^{-1}$)} & \colhead{(km s$^{-1}$)} & \colhead{} & \colhead{(K)} & \colhead{(km s$^{-1}$)} & \colhead{(K km s$^{-1}$)} & \colhead{(km s$^{-1}$)}
}
\startdata
HC$_{3}$N & & & & & & & & & & & & & & & & \\
$J=5-4$ & 45.490314 & 6.5 & 1.64 (4) & 3.35 (9) & 5.8 (2) & 32.8 & & 1.76 (5) & 3.50 (12) & 6.6 (3) & 18.7 & & 1.62 (6) & 2.22 (10) & 3.8 (2) & 49.0 \\
$J=10-9$ & 90.979023 & 24.0 & 2.58 (12) & 4.1 (2) & 11.3 (8) & 32.5 & & 3.15 (7) & 3.22 (8) & 10.8 (4) & 17.7 & & 1.64 (3) & 2.39 (5) & 4.16 (11) & 49.5 \\
H$^{13}$CCCN & & & & & & & & & & & & & & & & \\
$J=5-4$ & 44.084172 & 6.4 & 0.037 (7) & 3.0 (7) & 0.12 (4) & 33.0 & &  $< 0.036$ & ... & ... & ... & & 0.036 (6) & 2.7 (5) & 0.10 (3) & 49.8 \\
HC$^{13}$CCN & & & & & & & & & & & & & & & & \\
$J=5-4$ & 45.2973345 & 6.5 & $< 0.027$ & ... & ... & ... & & 0.050 (12) & 2.8 (8) & 0.15 (6) & 18.9 & & 0.051 (8) & 3.0 (6) & 0.16 (4) & 48.7 \\
HCC$^{13}$CN & & & & & & & & & & & & & & & & \\
$J=5-4$ & 45.3017069 & 6.5 & 0.042 (7) & 3.4 (7) & 0.15 (4) & 32.5 & & 0.090 (13) & 3.7 (6) & 0.36 (8) & 19.3 & & 0.073 (13) & 1.7 (4) & 0.13 (4) & 48.8 \\
DC$_{3}$N & & & & & & & & & & & & & & & & \\
$J=12-11$ & 101.314818 & 31.6 & $< 0.013$ & ... & ... & ... & & 0.034 (6) & 4.3 (9) & 0.16 (4) & 18.2 & & 0.014 (3) & 3.2 (7) & 0.049 (13) & 49.2 \\
\enddata
\tablecomments{Numbers in the parentheses are the standard deviation of the Gaussian fit, expressed in units of the last significant digits. For example, 1.64 (4) means $1.64 \pm 0.04$. The upper limits correspond to the $3 \sigma$ limits.}
\tablenotetext{a}{Taken from the Cologne Database for Molecular Spectroscopy \citep[CDMS;][]{2005JMoSt...742...215}.}
\tablenotetext{b}{The errors are 0.85 km s$^{-1}$, which corresponds to the velocity resolution (Section \ref{sec:obsNRO}).}
\end{deluxetable}

\begin{figure}
\figurenum{2}
\plotone{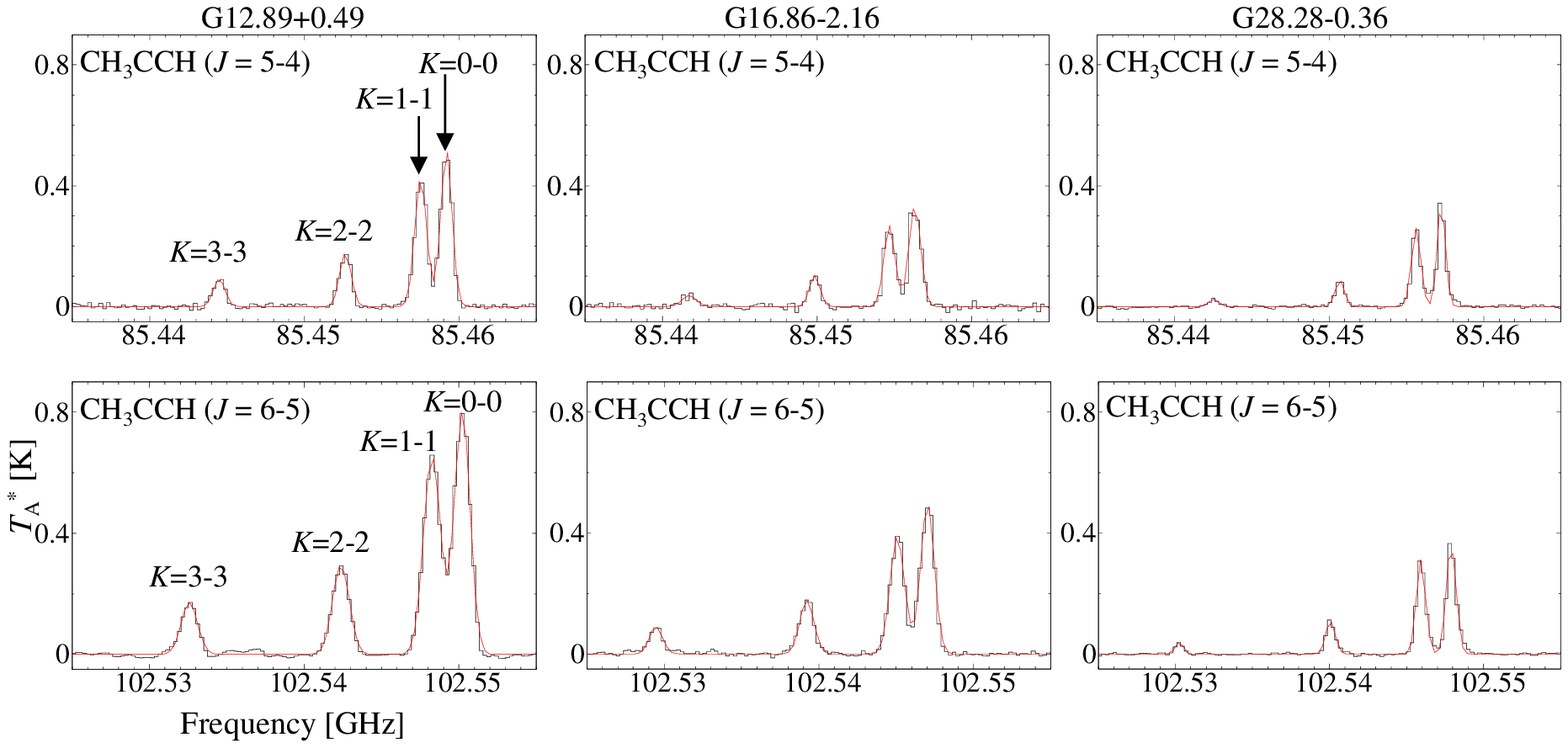}
\caption{Spectra of CH$_{3}$CCH in the three sources. The red lines show the Gaussian fitting results.\label{fig:f2}}
\end{figure}

Figure \ref{fig:f2} shows the spectra of CH$_{3}$CCH obtained with the Nobeyama 45-m radio telescope.
Its $J=5-4$ and $6-5$ $K$-ladder lines ($K=0-0$, $1-1$, $2-2$, and $3-3$) were detected from the three sources with a S/N ratio above 4. 
We fitted the spectra with four-component Gaussian profiles.
We fixed the centroid velocities to be the systemic velocities for each source (Table \ref{tab:source}).
The spectral line parameters are summarized in Table \ref{tab:resCH3CCH}.
There is no presence of wing emission in the CH$_{3}$CCH spectra and the lines are well fitted with the Gaussian profile.

\floattable
\rotate
\begin{deluxetable}{lllccccccccccc}
\tabletypesize{\scriptsize}
\tablecaption{Spectral line parameters of CH$_{3}$CCH in the three sources with the Nobeyama 45-m telescope \label{tab:resCH3CCH}}
\tablewidth{0pt}
\tablehead{
\colhead{} & \colhead{} & \colhead{}  & \multicolumn{3}{c}{G12.89+0.49} & \colhead{} & \multicolumn{3}{c}{G16.86$-$2.16} & \colhead{} & \multicolumn{3}{c}{G28.28$-$0.36} \\
\cline{4-6}\cline{8-10}\cline{12-14}
\colhead{Transition} & \colhead{Frequency\tablenotemark{a}} & \colhead{$E_{\rm {u}}/k$} & \colhead{$T_{\rm {mb}}$} & \colhead{FWHM} & \colhead{$\int T_{\mathrm {mb}}dv$} & \colhead{} & \colhead{$T_{\rm {mb}}$} & \colhead{FWHM} & \colhead{$\int T_{\mathrm {mb}}dv$} & \colhead{} & \colhead{$T_{\rm {mb}}$} & \colhead{FWHM} & \colhead{$\int T_{\mathrm {mb}}dv$} \\
\colhead{} & \colhead{(GHz)} & \colhead{(K)} & \colhead{(K)} & \colhead{(km s$^{-1}$)} & \colhead{(K km s$^{-1}$)} & \colhead{} & \colhead{(K)} & \colhead{(km s$^{-1}$)} & \colhead{(K km s$^{-1}$)} & \colhead{} & \colhead{(K)} & \colhead{(km s$^{-1}$)} & \colhead{(K km s$^{-1}$)}
}
\startdata
$J=5-4$ & & & & & & & & & & & & & \\
$K=0-0$ & 85.4573003 & 12.3 & 0.958 (14) & 3.03 (5) & 3.09 (6) & & 0.616 (18) & 3.18 (9) & 2.08 (8) & & 0.638 (14) & 2.19 (5) & 1.49 (4) \\
$K=1-1$ & 85.4556667 & 19.5 & 0.785 (14) & 3.20 (6) & 2.67 (7) & & 0.496 (18) & 3.02 (11) & 1.59 (8) & & 0.477 (13) & 2.46 (7) & 1.25 (5) \\
$K=2-2$ & 85.4507663 & 41.2 & 0.324 (14) & 3.08 (13) & 1.06 (6) & & 0.190 (19) & 2.8 (3) & 0.57 (8) & & 0.160 (13) & 2.44 (19) & 0.42 (5) \\
$K=3-3$ & 85.4426012 & 77.3 & 0.168 (14) & 3.2 (3) & 0.56 (7) & & 0.065 (16) & 3.9 (9) & 0.27 (9) & & 0.044 (13) & 2.6 (7) & 0.12 (5) \\
$J=6-5$ & & & & & & & & & & & & & \\
$K=0-0$ & 102.5479844 & 17.2 & 1.741 (13) & 3.51 (3) & 6.50 (8) & & 1.079 (14) & 3.07 (5) & 3.53 (7) & & 0.791 (12) & 2.16 (4) & 1.82 (4) \\
$K=1-1$ & 102.5460242 & 24.5 & 1.412 (13) & 3.81 (4) & 5.72 (8) & & 0.845 (13) & 3.26 (6) & 2.93 (7) & & 0.676 (12) & 2.09 (4) & 1.50 (4) \\
$K=2-2$ & 102.5401446 & 46.1 & 0.637 (13) & 3.75 (9) & 2.54 (8) & & 0.385 (13) & 3.19 (13) & 1.31 (7) & & 0.239 (12) & 2.09 (12) & 0.53 (4) \\
$K=3-3$ & 102.5303476 & 82.3 & 0.374 (13) & 3.61 (14) & 1.44 (8) & & 0.192 (14) & 3.0 (2) & 0.61 (7) & & 0.086 (13) & 1.8 (3) & 0.16 (4) \\
\enddata
\tablecomments{Numbers in the parentheses are the standard deviation of the Gaussian fit, expressed in units of the last significant digits.}
\tablenotetext{a}{Taken from the Cologne Database for Molecular Spectroscopy \citep[CDMS;][]{2005JMoSt...742...215}.}
\end{deluxetable}

Nine thermal CH$_{3}$OH lines were detected from G12.89+0.49, and eight lines, except for $6_{-2, 5}-7_{-1, 7}$ $E$ transition, were detected from the other two sources with a S/N ratio above 4 as shown in Figure \ref{fig:f3}.
We fitted the spectra with Gaussian profiles and the spectral line parameters are summarized in Table \ref{tab:resCH3OH}.
For the four lines shown in the top panels of Figure \ref{fig:f3}, we applied the four-component Gaussian fitting.
In the same way as for CH$_{3}$CCH, the centroid velocities were fixed to be the systemic velocities of each source (Table \ref{tab:source}).
The wing emission for red and blue velocity components are seen in G16.86$-$2.16, while only the red component is found in G12.89+0.49.
In G28.28$-$0.36, no wing emission is seen, which may be due to their lower line intensities.

\floattable
\rotate
\begin{deluxetable}{lllcccccccccccccc}
\tabletypesize{\scriptsize}
\tablecaption{Spectral line parameters of CH$_{3}$OH in the three sources with the Nobeyama 45-m telescope \label{tab:resCH3OH}}
\tablewidth{0pt}
\tablehead{
\colhead{} & \colhead{} & \colhead{}  & \multicolumn{4}{c}{G12.89+0.49} & \colhead{} & \multicolumn{4}{c}{G16.86$-$2.16} & \colhead{} & \multicolumn{4}{c}{G28.28$-$0.36} \\
\cline{4-7}\cline{9-12}\cline{14-17}
\colhead{Transition} & \colhead{Frequency\tablenotemark{a}} & \colhead{$E_{\rm {u}}/k$} & \colhead{$T_{\rm {mb}}$} & \colhead{FWHM} & \colhead{$\int T_{\mathrm {mb}}dv$} & \colhead{$V_{\rm{LSR}}$\tablenotemark{b}} & \colhead{} & \colhead{$T_{\rm {mb}}$} & \colhead{FWHM} & \colhead{$\int T_{\mathrm {mb}}dv$} & \colhead{$V_{\rm{LSR}}$\tablenotemark{b}} & \colhead{} & \colhead{$T_{\rm {mb}}$} & \colhead{FWHM} & \colhead{$\int T_{\mathrm {mb}}dv$} & \colhead{$V_{\rm{LSR}}$\tablenotemark{b}} \\
\colhead{} & \colhead{(GHz)} & \colhead{(K)} & \colhead{(K)} & \colhead{(km s$^{-1}$)} & \colhead{(K km s$^{-1}$)} & \colhead{(km s$^{-1}$)} & \colhead{} & \colhead{(K)} & \colhead{(km s$^{-1}$)} & \colhead{(K km s$^{-1}$)} & \colhead{(km s$^{-1}$)} & \colhead{} & \colhead{(K)} & \colhead{(km s$^{-1}$)} & \colhead{(K km s$^{-1}$)} & \colhead{(km s$^{-1}$)}
}
\startdata
$5_{-1,5}-4_{0,4}$ $E$ & 84.521169 & 40.4 & 1.97 (8) & 3.44 (14) & 7.2 (4) &  32.5 & & 1.20 (6) & 4.9 (2) & 6.2 (4) & 17.9 & & 0.222 (8) & 2.79 (10) & 0.66 (3) & 48.9 \\
$6_{-2, 5}-7_{-1, 7}$ $E$ & 85.568084 & 74.7 & 0.206 (7) & 5.06 (17) & 1.11 (5) & 32.9 & & $<0.018$ & ... & ... & ... & & $<0.011$ & ... & ... & ... \\
$8_{0, 8}-7_{1, 7}$ $A^{+}$ & 95.169463 & 83.5 & 1.82 (7) & 3.70 (15) & 7.2 (4) & 32.4 & & 0.96 (4) & 2.92 (13) & 2.99 (18) & 17.7 & & 0.045 (4) & 3.0 (3) & 0.14 (2) & 49.3 \\
$2_{1, 2}-1_{1, 1}$ $A^{+}$ & 95.914309 & 21.4 & 0.664 (13) & 4.39 (10) & 3.10 (9) & 32.7 & & 0.141 (4) & 5.55 (19) & 0.83 (4) & 18.0 & & 0.050 (4) & 2.8 (3) & 0.149 (19) & 48.9 \\
$2_{-1, 2}-1_{-1, 1}$ $E$ & 96.739362 & 12.5 & 1.81 (2) & 4.09 (6) & 7.87 (15) & $-$ & & 1.63 (2) & 5.31 (9) & 9.2 (2) & $-$ & & 1.06 (3) & 2.80 (10) & 3.16 (14) & $-$ \\
$2_{0, 2}-1_{0, 1}$ $A^{+}$ & 96.741375 & 7.0 & 2.39 (2) & 3.60 (4) & 9.14 (14) & $-$ & & 2.05 (2) & 4.50 (7) & 9.81 (19) & $-$ & & 1.42 (3) & 3.02 (7) & 4.57 (15) & $-$ \\
$2_{0 , 2}-1_{0 , 1}$ $E$ & 96.744550 & 20.1 & 0.99 (2) & 4.40 (11) & 4.66 (15) & $-$ & & 0.485 (18) & 7.8 (4) & 4.0 (2) & $-$ & & 0.23 (3) & 3.5 (5) & 0.87 (16) & $-$ \\
$2_{1 , 1}-1_{1 , 0}$ $E$ & 96.755511 & 28.0 & 0.64 (2) & 4.47 (17) & 3.06 (15) & $-$ & & 0.16 (2) & 5.2 (8) & 0.87 (18) & $-$ & & 0.08 (3) & 2.6 (12) & 0.22 (14) & $-$ \\
$2_{1, 1}-1_{1, 0}$ $A^{-}$ & 97.582804 & 21.6 & 0.702 (12) & 4.18 (8) & 3.12 (8) & 32.7 & & 0.172 (7) & 5.3 (2) & 0.97 (6) & 17.6 & & 0.065 (6) & 2.7 (3) & 0.18 (2) & 48.8 \\
\enddata
\tablecomments{Numbers in the parentheses are the standard deviation of the Gaussian fit, expressed in units of the last significant digits. The upper limits correspond to the $3 \sigma$ limits.}
\tablenotetext{a}{Taken from the Cologne Database for Molecular Spectroscopy \citep[CDMS;][]{2005JMoSt...742...215}.}
\tablenotetext{b}{The errors are 0.85 km s$^{-1}$, which corresponds to the velocity resolution (Section \ref{sec:obsNRO}).}
\end{deluxetable}

\begin{figure}
\figurenum{4}
\plotone{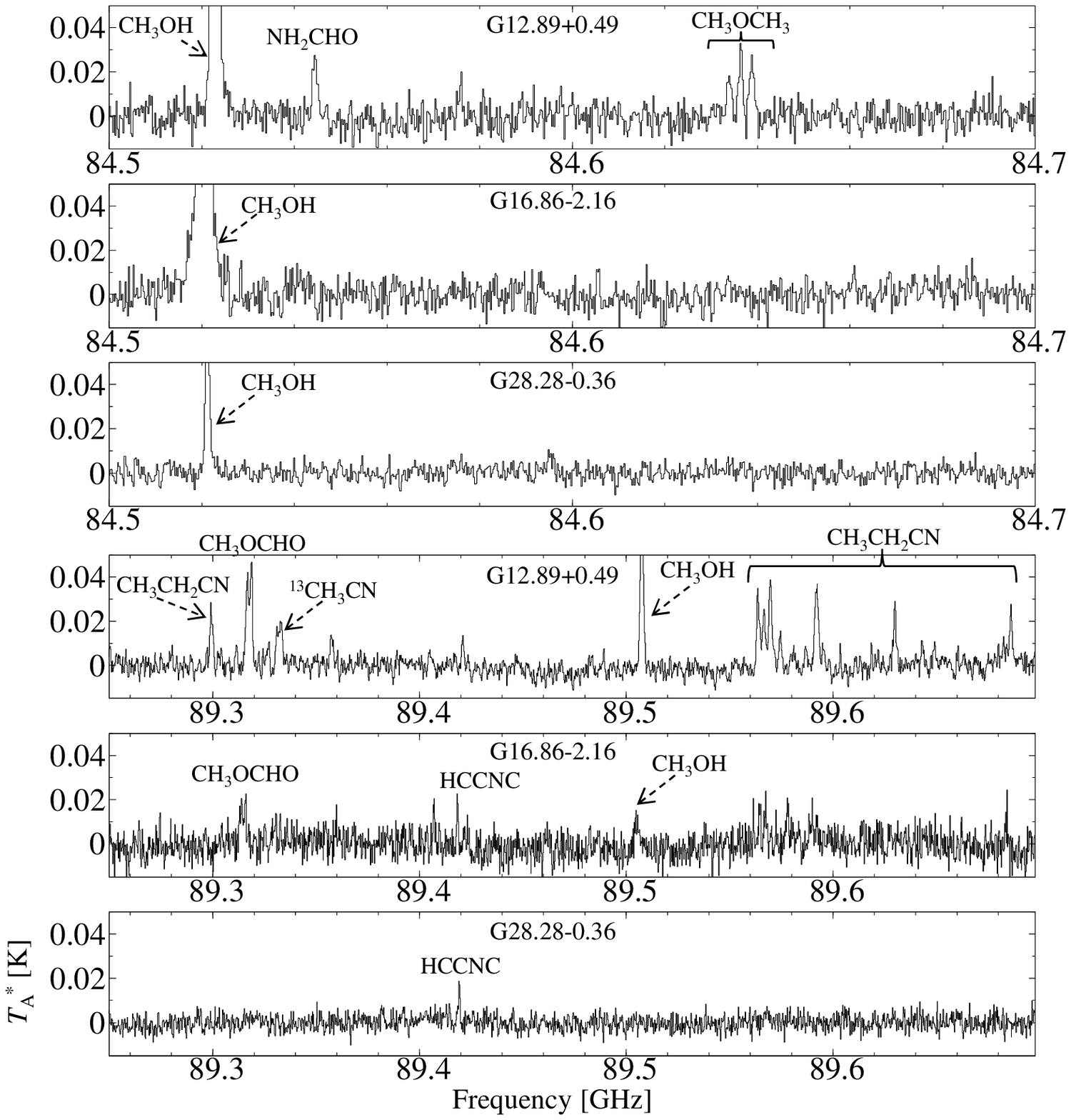}
\caption{Spectra of the complex organic molecules in the 84.5$-$84.7 and 89.3$-$89.7 GHz bands in the three sources obtained with the Nobeyama 45-m telescope. \label{fig:f6}}
\end{figure}

\begin{figure}
\figurenum{5}
\plotone{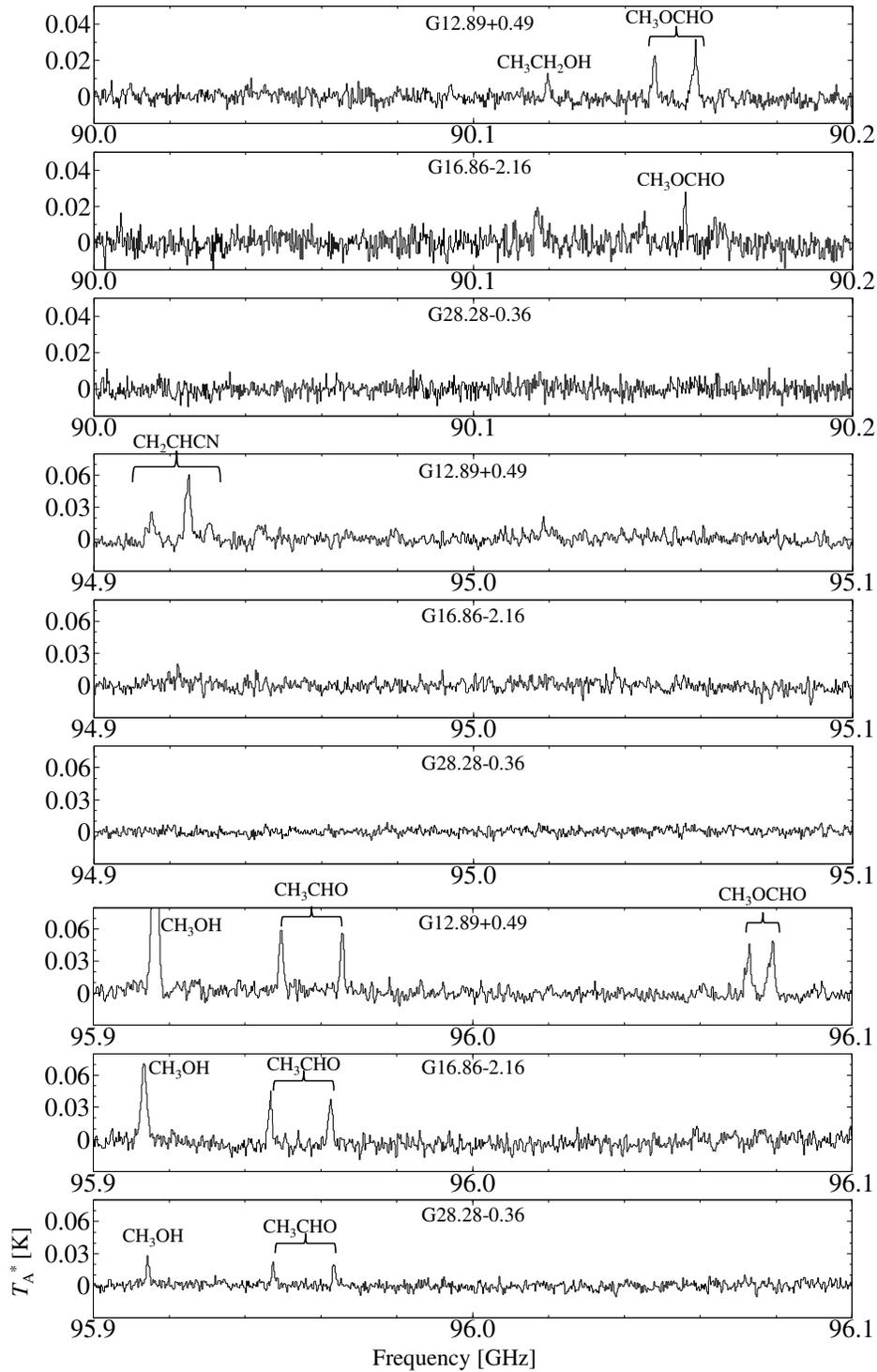}
\caption{Spectra of the complex organic molecules in the 90.0$-$90.2, 94.9$-$95.1, and 95.9$-$96.1 GHz bands in the three sources obtained with the Nobeyama 45-m telescope. \label{fig:f7}}
\end{figure}

Figures \ref{fig:f6} and \ref{fig:f7} show the spectra in the frequency bands covered with the TZ receiver.
Several lines from COMs have been detected with a S/N ratio above 4.
An analysis of the COMs requires the simultaneous fitting of multiple species to avoid blending effects with other lines. 
In our case, detected COMs have similar excitation-energy and such fitting cannot degenerate excitation temperatures and column densities. 
Therefore, we cannot derive their rotational temperatures and column densities accurately and we will not discuss COMs in the rest of this paper.
We summarize the detection and non-detection of COMs in each source in Table \ref{tab:resCOMs}.
G12.89+0.49 is the most line-rich source with strong peak intensities, and both nitrogen-bearing COMs and oxygen-bearing COMs have been detected.
On the other hand, only CH$_{3}$OH and CH$_{3}$CHO were detected  with weak peak intensities in G28.28$-$0.36.

\floattable
\begin{deluxetable}{lccc}
\tablecaption{Summary of detection of complex organic molecules in the three sources with the Nobeyama 45-m telescope \label{tab:resCOMs}}
\tablewidth{0pt}
\tablehead{
\colhead{Species} & \colhead{G12.89+0.49} & \colhead{G16.86$-$2.16} & \colhead{G28.28$-$0.36}
}
\startdata
CH$_{3}$OCHO & Y & Y & N \\
CH$_{3}$CH$_{2}$OH & Y & N & N \\
CH$_{3}$CHO & Y & Y & Y  \\
CH$_{3}$OCH$_{3}$ & Y & N & N  \\
NH$_{2}$CHO & Y & N & N \\
CH$_{2}$CHCN & Y & N & N \\
CH$_{3}$CH$_{2}$CN & Y & N & N \\
\enddata
\tablecomments{`Y' and `N' represent detection and non-detection with a S/N ratio above 4, respectively.}
\end{deluxetable}

\subsection{Observational Results with the ASTE 10-m Telescope} \label{sec:resASTE}

\begin{figure}
\figurenum{6}
\plotone{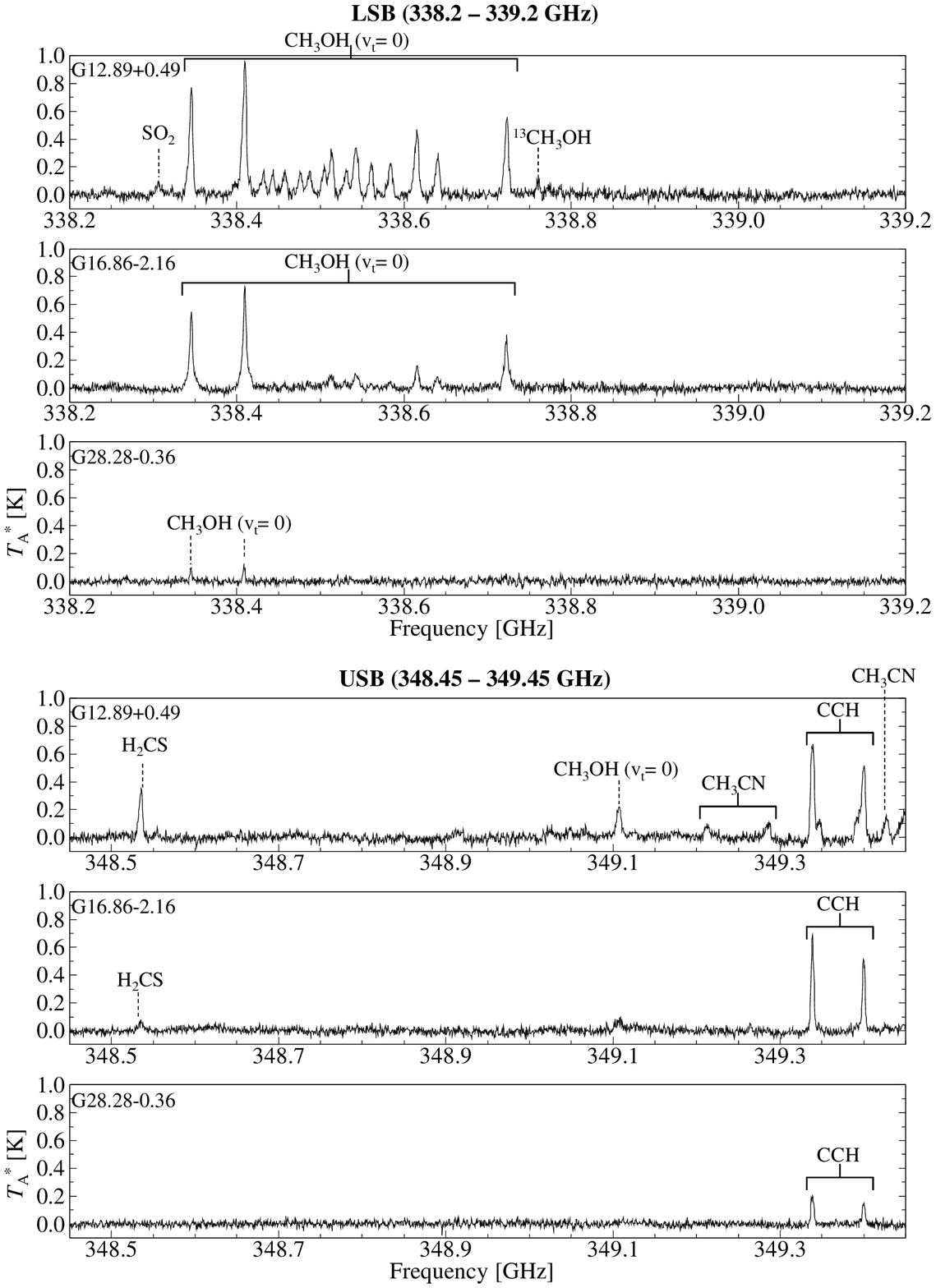}
\caption{Spectra in the 338.2$-$339.2 and 348.45$-$349.45 GHz bands toward the three sources with the ASTE 10-m telescope.\label{fig:f5}}
\end{figure}

Figure \ref{fig:f5} shows the spectra in the 338.2$-$339.2 and 348.45$-$349.45 GHz bands obtained with the ASTE 10-m telescope.
Table \ref{tab:resASTE} summarizes the spectral line parameters obtained from the Gaussian fitting.
The detection limit was set at a S/N ratio above 4.
The $V_{\rm {LSR}}$ values are consistent with the systemic velocities of each source (Table \ref{tab:source})\footnote{The $V_{\rm {LSR}}$ values of molecular emission lines are shifted by $\sim 0.2$ km s$^{-1}$ due to bug in ASTE Newstar. Results and discussions in this paper are not affected by this bug.}.
Several high-excitation energy ($E_{\rm {u}}/k > 200$ K) lines from CH$_{3}$OH and CH$_{3}$CN were detected from G12.89+0.49. 
On the other hand, only two lower-excitation energy ($E_{\rm {u}}/k < 75$ K) lines of CH$_{3}$OH and CCH were detected from G28.28$-$0.36.
The line density in G16.86$-$2.16 is between those in G12.89+0.49 and G28.28$-$0.36. 
The relatively high-excitation energy line of CH$_{3}$OH ($7_{4, 3}- 6_{4, 2}$ $A^{+}$ transition; $E_{\rm {u}}/k = 145.3$ K) was detected in G16.86$-$2.16.
The results suggest that hot gas exists in G12.89+0.49 and G16.86$-$2.16, which is supported by the detection of the metastable inversion transition NH$_{3}$ lines with the very-high-excitation energies, ($J, K$) = (8, 8) at 26.51898 GHz ($E_{\rm {u}}/k = 686 $ K) and ($J, K$) = (9, 9) at 27.47794 GHz ($E_{\rm {u}}/k = 852 $ K), with the GBT \citep{2017ApJ...844...68T}.

In G12.89+0.49 and G16.86$-$2.16, the line profiles of CH$_{3}$OH show wing emission suggestive of molecular outflow origins.
In addition, the line widths (FWHM) of CH$_{3}$OH in these two sources ($\sim 5-6$ km s$^{-1}$) are larger than those in G28.28$-$0.36 ($\sim 2-3$ km s$^{-1}$).
This may suggest that CH$_{3}$OH in G28.28$-$0.36 exists in less turbulent regions, as it is discussed in more detail in Section \ref{sec:discuss1}.

\floattable
\rotate
\begin{deluxetable}{lllcccccccccccccc}
\tabletypesize{\scriptsize}
\tablecaption{Spectral line parameters in the three sources with the ASTE 10-m telescope \label{tab:resASTE}}
\tablewidth{0pt}
\tablehead{
\colhead{} & \colhead{} & \colhead{}  & \multicolumn{4}{c}{G12.89+0.49} & \colhead{} & \multicolumn{4}{c}{G16.86$-$2.16} & \colhead{} & \multicolumn{4}{c}{G28.28$-$0.36} \\
\cline{4-7}\cline{9-12}\cline{14-17}
\colhead{Species \&} & \colhead{Frequency\tablenotemark{a}} & \colhead{$E_{\rm {u}}/k$} & \colhead{$T_{\rm {mb}}$} & \colhead{FWHM} & \colhead{$\int T_{\mathrm {mb}}dv$\tablenotemark{b}} & \colhead{$V_{\rm{LSR}}$\tablenotemark{c}} & \colhead{} & \colhead{$T_{\rm {mb}}$} & \colhead{FWHM} & \colhead{$\int T_{\mathrm {mb}}dv$\tablenotemark{b}} & \colhead{$V_{\rm{LSR}}$\tablenotemark{c}} & \colhead{} & \colhead{$T_{\rm {mb}}$} & \colhead{FWHM} & \colhead{$\int T_{\mathrm {mb}}dv$\tablenotemark{b}} & \colhead{$V_{\rm{LSR}}$\tablenotemark{c}} \\
\colhead{Transition} & \colhead{(GHz)} & \colhead{(K)} & \colhead{(K)} & \colhead{(km s$^{-1}$)} & \colhead{(K km s$^{-1}$)} & \colhead{(km s$^{-1}$)} & \colhead{} & \colhead{(K)} & \colhead{(km s$^{-1}$)} & \colhead{(K km s$^{-1}$)} & \colhead{(km s$^{-1}$)} & \colhead{} & \colhead{(K)} & \colhead{(km s$^{-1}$)} & \colhead{(K km s$^{-1}$)} & \colhead{(km s$^{-1}$)}
}
\startdata
SO$_{2}$ & \multicolumn{16}{c}{} \\
$18_{4,14}-18_{3,15}$ & 338.305993 & 196.8 & 0.13 (4)\tablenotemark{d} & 5.9 (9) & 0.81 (9) & 33.1 & & $<0.09$ & ... & ... & ... & & $<0.09$ & ... & ...& ... \\
CH$_{3}$OH &  \multicolumn{16}{c}{} \\
$7_{-1, 7}- 6_{-1, 6}$ $E$ & 338.344628 & 70.5 & 1.23 (4) & 4.6 (9) & 5.99 (10) & 32.9 & & 0.82 (3) & 4.5 (9) & 3.94 (8) & 17.3 & & 0.17 (3) & 2.2 (8) & 0.40 (6) & 48.9 \\
$7_{0, 7}- 6_{0, 6}$ $A^{+}$ & 338.408681 & 65.0 & 1.50 (4) & 5.1 (9) & 8.06 (11) & 33.0 & & 1.07 (3) & 4.8 (9) & 5.42 (7) & 17.2 & & 0.19 (3) & 2.5 (8) & 0.49 (5) & 49.2 \\
$7_{-6, 1}- 6_{-6, 0}$ $E$ & 338.430933 & 253.9 & 0.23 (4) & 5.8 (9) & 1.40 (9) & 32.9 & & $<0.09$ & ... & ... & ... & & $<0.09$ & ... & ...& ... \\
$7_{6, 1}- 6_{6, 0}$ $A^{+}$ & 338.442344 & 258.7 & 0.26 (4) & 4.5 (9) & 1.22 (9) & 33.0 & & $<0.09$ & ... & ... & ... & & $<0.09$ & ... & ...& ... \\
$7_{-5, 2}- 6_{-5, 1}$ $E$ & 338.456499 & 189.0 & 0.26 (4) & 5.0 (9) & 1.38 (11) & 32.9 & & $<0.09$ & ... & ... & ... & & $<0.09$ & ... & ...& ... \\
$7_{5 , 3}- 6_{5 , 2}$ $E$ & 338.475290 & 201.1 & 0.25 (4) & 5.2 (9) & 1.42 (11) & 32.7 & & $<0.09$ & ... & ... & ... & & $<0.09$ & ... & ...& ... \\
$7_{5, 3}- 6_{5, 2}$ $A^{+}$ & 338.486337 & 202.9 & 0.25 (4) & 5.9 (9) & 1.55 (11) & 33.2 & & $<0.09$ & ... & ... & ... & & $<0.09$ & ... & ...& ... \\
$7_{-4, 4}- 6_{-4, 3}$ $E$ & 338.504099 & 152.9 & 0.30 (4) & 6.0 (9) & 1.91 (10) & 32.8 & & $<0.09$ & ... & ... & ... & & $<0.09$ & ... & ...& ... \\
$7_{4, 3}- 6_{4, 2}$ $A^{+}$ & 338.512627 & 145.3 & 0.51 (4) & 4.7 (9) & 2.56 (11) & 32.7 & & 0.14 (3) & 6.1 (9) & 0.92 (10) & 17.7 & & $<0.09$ & ... & ...& ... \\
$7_{4 , 3}- 6_{4 , 2}$ $E$ & 338.530249 & 161.0 & 0.27 (4) & 6.0 (9) & 1.72 (9) & 32.8 & & $<0.09$ & ... & ... & ... & & $<0.09$ & ... & ...& ... \\
$7_{3, 5}- 6_{3, 4}$ $A^{+}$ & 338.540795 & 114.8 & 0.56 (4) & 6.1 (9) & 3.64 (12) & 32.0 & & 0.15 (3) & 6.7 (9) & 1.10 (10) & 16.1 & & $<0.09$ & ... & ...& ... \\
$7_{-3, 5}- 6_{-3, 4}$ $E$ & 338.559928 & 127.7 & 0.36 (4) & 4.5 (9) & 1.73 (10) & 32.8 & & $<0.09$ & ... & ... & ... & & $<0.09$ & ... & ...& ... \\
$7_{3 , 4}- 6_{3 , 3}$ $E$ & 338.583195 & 112.7 & 0.36 (4) & 5.2 (9) & 2.00 (10) & 33.0 & & $<0.09$ & ... & ... & ... & & $<0.09$ & ... & ...& ... \\
$7_{1 , 6}- 6_{1 , 5}$ $E$ & 338.614999 & 86.1 & 0.73 (4) & 5.5 (9) & 4.28 (11) & 33.3 & & 0.26 (3) & 4.3 (9) & 1.20 (8) & 17.3 & & $<0.09$ & ... & ...& ... \\
$7_{2, 5}- 6_{2, 4}$ $A^{+}$ & 338.639939 & 102.7 & 0.45 (4) & 4.9 (9) & 2.36 (11) & 33.2 & & 0.12 (3) & 5.8 (9) & 0.72 (9) & 17.9 & & $<0.09$ & ... & ...& ... \\
$7_{-2, 6}- 6_{-2, 5}$ $E$ & 338.72294 & 90.9 & 0.89 (4) & 4.9 (9) & 4.64 (11) & 32.3 & & 0.51 (3) & 5.8 (9) & 3.10 (7) & 18.2 & & $<0.09$ & ... & ...& ... \\
$14_{1, 13}- 14_{0, 14}$ $A^{- +}$ & 349.10702 & 260.2 & 0.36 (4) & 6.0 (9) & 2.29 (12) & 33.2 & & $<0.09$ & ... & ... & ... & & $<0.09$ & ... & ...& ... \\
$^{13}$CH$_{3}$OH & \multicolumn{16}{c}{} \\
$13_{0, 13}- 12_{1, 12}$ $A^{+}$ & 338.759948 & 205.9 & 0.17 (4) & 4.5 (9) & 0.80 (6) & 33.0 & & $<0.09$ & ... & ... & ... & & $<0.09$ & ... & ...& ... \\
H$_{2}$CS & \multicolumn{16}{c}{} \\
$10_{1, 9}- 9_{1, 8}$ & 348.5343647 & 105.2 & 0.55 (4) & 4.5 (9) & 2.64 (11) & 32.9 & & 0.10 (3) & 6.2 (9) & 0.64 (6) & 17.4 & & $<0.09$ & ... & ...& ... \\
CCH &  \multicolumn{16}{c}{} \\
$N= 4- 3, J=9/2-7/2$ & 349.3374558 & 41.9 & 1.17 (4) & 4.2 (9) & 5.22 (11) & 33.7 & & 1.12 (3) & 3.2 (9) & 3.82 (8) & 16.8 & & 0.34 (3) & 3.4 (9) & 1.25 (7) & 48.1 \\
$N= 4- 3, J=7/2-5/2$ & 349.3992738 & 41.9 & 0.80 (4)\tablenotemark{e} & 3.8 (9) & 3.27 (11) & 32.1 & & 0.86 (3) & 3.3 (9) & 3.04 (8) & 17.1 & & 0.25 (3) & 3.4 (9) & 0.92 (7) & 48.3 \\
CH$_{3}$CN &  \multicolumn{16}{c}{} \\
$J_{K}=19_{6}-18_{6}$ & 349.2123106 & 424.7 & 0.14 (4) & 5.4 (9) & 0.81 (12) & 33.1 & & $<0.09$ & ... & ... & ... & & $<0.09$ & ... & ...& ... \\
$J_{K}=19_{5}-18_{5}$ & 349.2860057 & 346.2 & 0.16 (4) & 5.8 (9) & 0.98 (12) & 33.9 & & $<0.09$ & ... & ... & ... & & $<0.09$ & ... & ...& ... \\
$J_{K}=19_{4}-18_{4}$ & 349.3463428 & 282.0 & 0.20 (4)\tablenotemark{e} & 5.1 (9) & 1.07 (11) & 33.3 & & $<0.09$ & ... & ... & ... & & $<0.09$ & ... & ...& ... \\
$J_{K}=19_{2}-18_{2}$ & 349.4268497 & 196.3 & 0.23 (4) & 5.3 (9) & 1.32 (11) & 33.0 & & $<0.09$ & ... & ... & ... & & $<0.09$ & ... & ...& ... \\
\enddata
\tablecomments{Numbers in the parentheses are the standard deviation, expressed in units of the last significant digits. The upper limits correspond to the $3 \sigma$ limits.}
\tablenotetext{a}{Taken from the Cologne Database for Molecular Spectroscopy \citep[CDMS;][]{2005JMoSt...742...215}.}
\tablenotetext{b}{The errors were derived using the following formula; $\Delta T_{\rm {mb}} \sqrt{n} \Delta v$, where $\Delta T_{\rm {mb}}$, $n$, $\Delta v$ are the rms noise in the emission-free regions, the number of channels, and velocity resolution per channel, respectively.}
\tablenotetext{c}{The errors are 0.86 km s$^{-1}$, which corresponds to the velocity resolution (Section \ref{sec:obsASTE}).}
\tablenotetext{d}{Tentative detection with the signal-to-noise ratio above 3.}
\tablenotetext{e}{These lines are contaminated and fitting results are tentative.}
\end{deluxetable}

\section{Analyses} \label{sec:ana}

\subsection{Rotational Diagram Analysis} \label{sec:rotational}

We derived the rotational temperatures and beam-averaged column densities of HC$_{3}$N, CH$_{3}$CCH, and CH$_{3}$OH with the beam size of $18\arcsec$ in the three sources from the rotational diagram analysis, using the following formula \citep{1999ApJ...517..209G};
\begin{equation} \label{rd}
{\rm {ln}} \frac{3k \int T_{\mathrm {mb}}dv}{8\pi ^3 \nu S \mu ^2} = {\rm {ln}} \frac{N}{Q(T_{\rm {rot}})} - \frac{E_{\rm {u}}}{kT_{\rm {rot}}},
\end{equation}
where $k$ is the Boltzmann constant, $S$ is the line strength, $\mu$ is the permanent electric dipole moment, $N$ is the column density, and $Q(T_{\rm {rot}})$ is the partition function.
The permanent electric dipole moments are 3.7312, 0.784, and 0.899 D for HC$_{3}$N, CH$_{3}$CCH, and CH$_{3}$OH, respectively.
We used $\int T_{\mathrm {mb}}dv$ values summarized in Tables \ref{tab:resHC3N}, \ref{tab:resCH3CCH}, and \ref{tab:resCH3OH}.

\begin{figure}
\figurenum{7}
\plotone{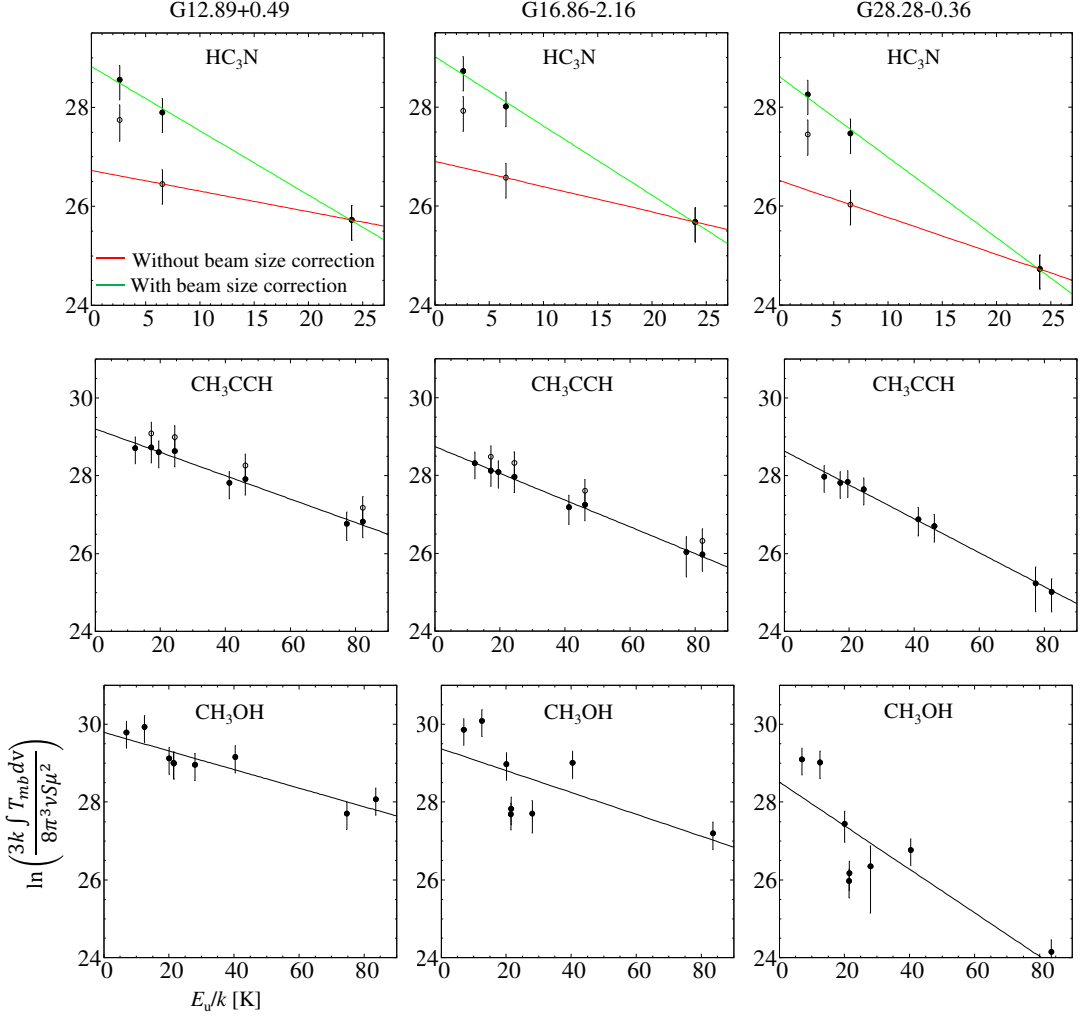}
\caption{Rotational diagram of HC$_{3}$N, CH$_{3}$CCH, and CH$_{3}$OH in the three sources. HC$_{3}$N (upper panels): The filled and open circles represent with and without beam size correction, respectively. The red line is the fitting results for the $J=5-4$ data without beam size correction and $J=10-9$ data. The green line shows the fitting results for all of the lines with the beam size correction. CH$_{3}$CCH (middle panels): The filled and open circles represent with and without correction of frequency dependence, respectively. The black lines show the fitting results with correction of frequency dependence. CH$_{3}$OH (lower panels): The black lines show the fitting results for all of the observed lines. \label{fig:f4}}
\end{figure}

Figure \ref{fig:f4} shows the fitting results of HC$_{3}$N, CH$_{3}$CCH, and CH$_{3}$OH in the three sources.
The errors include the Gaussian fitting errors, the uncertainties from the main beam efficiency of 10\%, the chopper-wheel method of 10\%, and pointing calibration error of 30\%.
The derived rotational temperatures and column densities are summarized in Table \ref{tab:rot}.

In the case of HC$_{3}$N, we added data of the $J=3-2$ transition obtained with the GBT \citep[Table 3 in][]{2017ApJ...844...68T}.
The beam sizes are different among the three transitions, $27\arcsec$, $37\arcsec$, and $18\arcsec$ for the $J=3-2$, $5-4$, and $10-9$, respectively.
Assuming a small beam filling factor, we multiplied the integrated intensities of the GBT data by ($\frac{27\arcsec}{18\arcsec}$)$^{2}$ and the 45 GHz band data by ($\frac{37\arcsec}{18\arcsec}$)$^{2}$ for the correction of the different beam sizes (filled circles in Figure \ref{fig:f4}). 
The red lines are the best fitting results for $J=5-4$ without beam size correction and $J=10-9$ transitions, and the green lines show the fitting results for the three transitions with beam size correction.
In the case of the fitting for the $J=5-4$ and $10-9$ transitions, the $1\sigma$ fluctuation in the integrated intensity causes unlikely large errors in the derived rotational temperature and column density (e.g., $T_{\rm {rot}} \sim 600$ K), and we do not include the errors in Table \ref{tab:rot}.
All of the three transitions are better fitted with beam size correction.
The fitting results imply that the spatial distribution of HC$_{3}$N is smaller than 18\arcsec, corresponding to $\sim 0.07-0.1$ pc radii at the source distances (Table \ref{tab:source}), like in the case for HC$_{5}$N \citep{2017ApJ...844...68T}.
\citet{2017ApJ...844...68T} derived the rotational temperatures of HC$_{5}$N with beam size correction to be $\sim 13-20$ K in the three sources.
The different rotational temperatures between HC$_{3}$N and HC$_{5}$N probably arise from the fact that we observed only lower excitation-energy lines ($E_{\rm {u}}/k < 24.0$ K), compared to HC$_{5}$N \citep[$E_{\rm {u}}/k \leq 100$ K;][]{2017ApJ...844...68T}, and the presence of complex excitation conditions.

In the case of CH$_{3}$CCH, we multiplied the integrated intensities of the $J=6-5$ transition lines by ($\frac{85.4 {\rm GHz}}{102.5 {\rm GHz}}$)$^{2}$ in order to correct the values for the $18\arcsec$ beam size at the 85 GHz band in G12.89+0.49 and G16.86$-$2.16, because data of the $J=6-5$ transition are systematically higher than those of the $J=5-4$ transition.
In Figure \ref{fig:f4}, open and filled circles represent the integrated intensities without and with correction of frequency dependence of the beam size, respectively.
The black lines show the fitting results with correction of frequency dependence.
All of the data can be fitted simultaneously, which means that the spatial distribution of CH$_{3}$CCH is smaller than $18\arcsec$.
This is the same case as for HC$_{3}$N and HC$_{5}$N.
The derived rotational temperatures are $33^{+20}_{-9}$, $29^{+15}_{-8}$, and $23^{+9}_{-6}$ K in G12.89+0.40, G16.86$-$2.16, and G28.28$-$0.36, respectively.

In the case of CH$_{3}$OH, we fitted the data using all the lines in the 85$-$98 GHz band summarized in Table \ref{tab:resCH3OH}, and the fitting results are shown by the black lines in Figure \ref{fig:f4}.
As is the case of CH$_{3}$CCH, we derived beam-averaged column densities a beam size of $18\arcsec$, correcting the frequency dependence of the beam size by multiplying the integrated intensities of the lines in the 95$-$98 GHz by ($\frac{85 {\rm GHz}}{\nu {\rm GHz}}$)$^{2}$, where $\nu$ is the frequency of each line.
The rotational temperatures are $42^{+33}_{-13}$, $36^{+18}_{-9}$, and $18^{+5}_{-3}$ K in G12.89+0.49, G16.86$-$2.16, and G28.28$-$0.36, respectively (Table \ref{tab:rot}).
\citet{2009MNRAS.394..323P} derived the rotational temperatures of CH$_{3}$OH to be $5\pm 0.8$, $6 \pm 0.1$, and $5 \pm 0.5$ K in G12.89+0.49, G16.86$-$2.16, and G28.28$-$0.36, respectively, from fitting only three or four line transitions, which was interpreted as CH$_{3}$OH being sub-thermally excited.
This does not seem to be the case when considering more CH$_{3}$OH transitions in the derivation of the rotational temperature.

\floattable
\begin{deluxetable}{ccccccccc}
\tablecaption{The rotational temperatures and beam-averaged column densities of HC$_{3}$N, CH$_{3}$CCH, and CH$_{3}$OH \label{tab:rot}}
\tablewidth{0pt}
\tablehead{
\colhead{} & \multicolumn{2}{c}{G12.89+0.49} & \colhead{} & \multicolumn{2}{c}{G16.86$-$2.16} & \colhead{} & \multicolumn{2}{c}{G28.28$-$0.36} \\
\cline{2-3}\cline{5-6}\cline{8-9}
\colhead{Species} & \colhead{$T_{\rm {rot}}$} & \colhead{$N$} & \colhead{} & \colhead{$T_{\rm {rot}}$} & \colhead{$N$} & \colhead{} & \colhead{$T_{\rm {rot}}$} & \colhead{$N$} \\
\colhead{} & \colhead{(K)} & \colhead{(cm$^{-2}$)} & \colhead{} & \colhead{(K)} & \colhead{(cm$^{-2}$)} & \colhead{} & \colhead{(K)} & \colhead{(cm$^{-2}$)}
}
\startdata
HC$_{3}$N & & & & & & & & \\
$J=5-4$ \& $10-9$ & $24$ & $4.4 \times 10^{13}$ & & $20$ & $4.3 \times 10^{13}$ & & $13.4$ & $2.0 \times 10^{13}$ \\
All & $7.7^{+1.6}_{-2.8}$ & ($1.4 \pm 0.3$)$\times 10^{14}$ & & $7.2^{+1.4}_{-2.4}$ & ($1.3 \pm 0.3$)$\times 10^{14}$ & & $6.2^{+1.1}_{-1.7}$ & ($7.6 \pm 2.1$)$\times 10^{13}$ \\
\multicolumn{9}{c}{} \\
CH$_{3}$CCH & $33^{+20}_{-9}$ & $1.0^{+0.11}_{-0.02} \times 10^{15}$ & & $29^{+15}_{-8}$ & $5.4^{+0.7}_{-0.8} \times 10^{14}$ & & $23^{+9}_{-6}$ & $3.7^{+0.5}_{-0.9} \times 10^{14}$ \\
\multicolumn{9}{c}{} \\
CH$_{3}$OH & $42^{+33}_{-13}$ & $2.9^{+1.9}_{-1.3} \times 10^{15}$ & & $36^{+18}_{-9}$ & $1.5^{+0.9}_{-0.7} \times 10^{15}$ & & $18^{+5}_{-3}$ & $2.3^{+1.6}_{-1.1} \times 10^{14}$ \\
\enddata
\tablecomments{The errors represent the standard deviation.}
\end{deluxetable}

\subsection{Column Densities of the D and $^{13}$C Isotopologues of HC$_{3}$N}

We derived the column densities of the D and $^{13}$C isotopologues of HC$_{3}$N in the three sources assuming the LTE condition \citep{1999ApJ...517..209G}.
We used the following formulae;

\begin{equation} \label{tau}
\tau = - {\mathrm {ln}} \left[1- \frac{T_{\rm {mb}} }{J(T_{\rm {ex}}) - J(T_{\rm {bg}})} \right]
\end{equation}
where
\begin{equation} \label{tem}
J(T) = \frac{h\nu}{k}\Bigl\{\exp\Bigl(\frac{h\nu}{kT}\Bigr) -1\Bigr\} ^{-1},
\end{equation}  
and
\begin{equation} \label{col}
N = \tau \frac{3h\Delta v}{8\pi ^3}\sqrt{\frac{\pi}{4\mathrm {ln}2}}Q\frac{1}{\mu ^2}\frac{1}{J_{\rm {lower}}+1}\exp\Bigl(\frac{E_{\rm {lower}}}{kT_{\rm {ex}}}\Bigr)\Bigl\{1-\exp\Bigl(-\frac{h\nu }{kT_{\rm {ex}}}\Bigr)\Bigr\} ^{-1}.
\end{equation} 
In Equation (\ref{tau}), $\tau$ denotes the optical depth, and $T_{\rm {mb}}$ the peak intensities summarized in Table \ref{tab:resHC3N}.
$T_{\rm{ex}}$ and $T_{\rm {bg}}$ are the excitation temperature and the cosmic microwave background temperature ($\simeq 2.73$ K).
We used the rotational temperatures of HC$_{3}$N summarized in Table \ref{tab:rot} as the excitation temperatures.
We calculated two cases of excitation temperatures (Table \ref{tab:rot}) for each source.
$J$($T$) in Equation (\ref{tem}) is the effective temperature equivalent to that in the Rayleigh-Jeans law.
In Equation (\ref{col}), {\it N} is the column density,  $\Delta v$ is the line width (FWHM, Table \ref{tab:resHC3N}), $Q$ is the partition function, $\mu$ is the permanent electric dipole moment, and $E_{\rm {lower}}$ is the energy of the lower rotational energy level. 
The electric dipole moments are 3.7408 D and 3.73172 D for DC$_{3}$N and the three $^{13}$C isotopologues, respectively.
The derived column densities and the D/H and $^{12}$C/$^{13}$C ratios are summarized in Table \ref{tab:LTE}.

\floattable
\begin{deluxetable}{ccccccccc}
\tablecaption{The column densities of the D and $^{13}$C isotopologues of HC$_{3}$N \label{tab:LTE}}
\tablewidth{0pt}
\tablehead{
\colhead{} & \multicolumn{2}{c}{G12.89+0.49} & \colhead{} & \multicolumn{2}{c}{G16.86$-$2.16} & \colhead{} & \multicolumn{2}{c}{G28.28$-$0.36} \\
\cline{2-3}\cline{5-6}\cline{8-9}
\colhead{Species} & \colhead{$N$} & \colhead{D/H,} & \colhead{} & \colhead{$N$} & \colhead{D/H,} & \colhead{} & \colhead{$N$} & \colhead{D/H,} \\
\colhead{} & \colhead{(cm$^{-2}$)} & \colhead{$^{12}$C/$^{13}$C} & \colhead{} & \colhead{(cm$^{-2}$)} & \colhead{$^{12}$C/$^{13}$C} & \colhead{} & \colhead{(cm$^{-2}$)} & \colhead{$^{12}$C/$^{13}$C}
}
\startdata
& \multicolumn{2}{c}{$T_{\rm {ex}} = 24$ K (fixed)} & & \multicolumn{2}{c}{$T_{\rm {ex}} = 20$ K (fixed)} & & \multicolumn{2}{c}{$T_{\rm {ex}} = 13.4$ K (fixed)} \\
\cline{2-3}\cline{5-6}\cline{8-9}
DC$_{3}$N & $< 6.2 \times 10^{10}$ & $< 0.0014$ & & ($5.5 \pm 1.5$)$\times 10^{11}$ &  $0.013 \pm 0.003$ & & ($3.7 \pm 1.0$)$\times 10^{11}$ & $0.018 \pm 0.005$ \\
H$^{13}$CCCN & ($5.8 \pm 1.7$)$\times 10^{11}$ & $76 \pm 23$ & & $<2.6 \times 10^{11}$ & $<165$ & & ($6.8 \pm 1.8$)$\times 10^{11}$ & $30 \pm 8$ \\  
HC$^{13}$CCN & $<1.9 \times 10^{11}$ & $< 237$ & & ($7.6 \pm 2.8$)$\times 10^{11}$ & $57 \pm 21$ & & ($1.0 \pm 0.3$)$\times 10^{12}$ & $20 \pm 5$ \\
HCC$^{13}$CN & ($7.1 \pm 1.9$)$\times 10^{11}$ & $63 \pm 16$ & & ($1.8 \pm 0.4$)$\times 10^{12}$ & $24 \pm 5$ & & ($8.2 \pm 2.3$)$\times 10^{11}$ & $25 \pm 7$ \\
\multicolumn{9}{c}{} \\
& \multicolumn{2}{c}{$T_{\rm {ex}} = 7.7$ K (fixed)} & & \multicolumn{2}{c}{$T_{\rm {ex}} = 7.2$ K (fixed)} & & \multicolumn{2}{c}{$T_{\rm {ex}} = 6.2$ K (fixed)} \\
\cline{2-3}\cline{5-6}\cline{8-9}
DC$_{3}$N & $< 5.4 \times 10^{11}$ & $< 0.0046$ & & ($4.9 \pm 1.4$) $\times 10^{12}$ & $0.037 \pm 0.010$ & & ($3.3 \pm 0.9$)$\times 10^{12}$ & $0.043 \pm 0.012$ \\
H$^{13}$CCCN & ($2.5 \pm 0.8$)$\times 10^{12}$ & $46 \pm 14$ & & $<1.2 \times 10^{12}$ & $<114$ & & ($3.1 \pm 0.8$)$\times 10^{12}$ & $24 \pm 7$ \\
HC$^{13}$CCN & $<8.2 \times 10^{11}$ & $<142$ & & ($3.4 \pm 1.3$)$\times 10^{12}$ & $38 \pm 14$ & & ($4.7 \pm 1.2$)$\times 10^{12}$ & $16 \pm 4$ \\
HCC$^{13}$CN & ($3.1 \pm 0.8$)$\times 10^{12}$ & $37 \pm 10$ & & ($6.0 \pm 1.4$)$\times 10^{12}$ & $22 \pm 5$ & & ($3.9 \pm 1.1$)$\times 10^{12}$ & $19 \pm 5$ \\ 
\enddata
\tablecomments{The errors represent the standard deviation. The upper limits were derived using the $1\sigma$ noise level.}
\end{deluxetable}

\section{Discussion} \label{sec:discuss}

In this section, we will discuss comparisons of the chemical composition among MYSOs using CH$_{3}$OH, CH$_{3}$CCH, and HC$_{5}$N, all of which are considered to be in the lukewarm envelopes (Table \ref{tab:rot}).
In the following discussion, we assume that the spatial distributions of CH$_{3}$OH, CH$_{3}$CCH, and HC$_{5}$N are similar to each other.
This assumption is based on the following results;
the source sizes of both HC$_{5}$N and CH$_{3}$CCH seem to be smaller than $18\arcsec$ (see \citet{2017ApJ...844...68T} and Section \ref{sec:rotational}), and the rotational temperatures of CH$_{3}$OH are comparable with those of CH$_{3}$CCH in our three sources.
Nitrogen-bearing COMs are usually originated from hot cores \citep[e.g.,][]{2007A&A...465..913B}, and thus we do not include them in discussion.
We will also discuss the relationship between the line width and excitation energy of rotational lines of CH$_{3}$OH.

\subsection{Comparisons of Fractional Abundances in the High-Mass Star-Forming Regions}  \label{sec:discuss2}

\begin{figure}
\figurenum{8}
\plotone{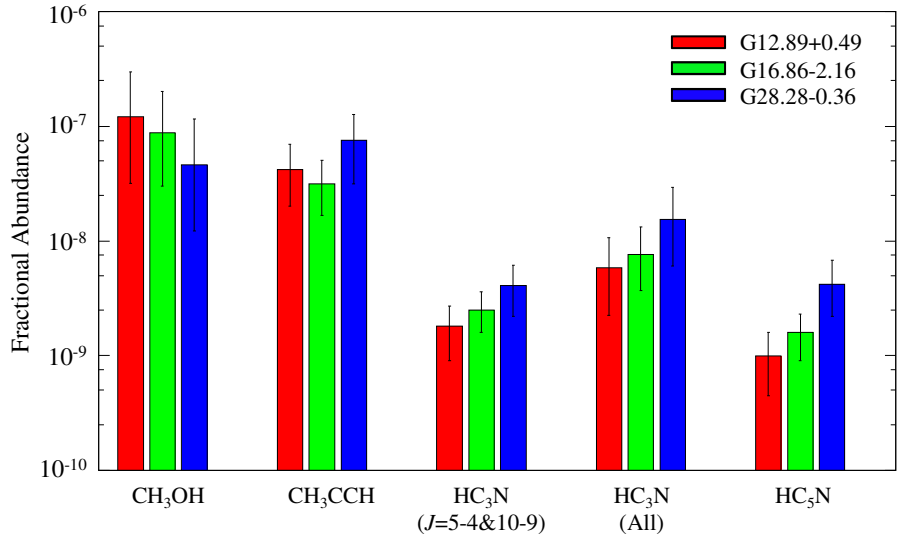}
\caption{Fractional abundances of CH$_{3}$OH, CH$_{3}$CCH, HC$_{3}$N, and HC$_{5}$N in the high-mass star-forming regions. The HC$_{5}$N values are taken from \citet{2017ApJ...844...68T}. \label{fig:f9}}
\end{figure}

We derived the fractional abundances of HC$_{3}$N, CH$_{3}$OH, and CH$_{3}$CCH, $X$($a$) = $N$($a$)/$N$(H$_{2}$), in the three sources.
The H$_{2}$ column densities, $N$(H$_{2}$), in the three sources were derived by \citet{2017ApJ...844...68T}.
The $N$(H$_{2}$) values and the fractional abundances of HC$_{3}$N, CH$_{3}$OH, CH$_{3}$CCH, and HC$_{5}$N derived by \citet{2017ApJ...844...68T} are summarized in Table \ref{tab:frac}. 
The derived $X$(CH$_{3}$OH) values are $1.2^{+1.8}_{-0.9} \times 10^{-7}$, $8.8^{+11}_{-5.8} \times 10^{-8}$, and $4.6^{+7.0}_{-3.4} \times 10^{-8}$ in G12.89+0.49, G16.86$-$2.16, and G28.28$-$0.36, respectively.
The $X$(CH$_{3}$CCH) values are derived to be $4.2^{+2.8}_{-2.2} \times 10^{-8}$, $3.2^{+1.9}_{-1.5} \times 10^{-8}$, and $7.6^{+5.2}_{-4.4} \times 10^{-8}$ in G12.89+0.49, G16.86$-$2.16, and G28.28$-$0.36, respectively.
We derive the $X$(HC$_{3}$N) values for two cases, fitting results of two transitions and all transitions (Table \ref{tab:rot}).
Figure \ref{fig:f9} shows their fractional abundances in the three sources.
All of the column densities including the molecular hydrogen were derived as beam-averaged values for a beam size of $18\arcsec$.
More accurate column densities and fractional abundances can be obtained in the future via interferometric observations that resolve the size of the sources.

Although our sample is small, there seems to be an anti-correlation between $X$(CH$_{3}$OH) and $X$(HC$_{5}$N).
G12.89+0.49 shows the highest $X$(CH$_{3}$OH) value and the lowest $X$(HC$_{5}$N) value among the three sources, while G28.28$-$0.36 shows the lowest $X$(CH$_{3}$OH) value and the highest $X$(HC$_{5}$N) value.
We further discuss the relationship between HC$_{5}$N and CH$_{3}$OH in Section \ref{sec:discuss3}.
The $X$(HC$_{3}$N) has a positive correlation with $X$(HC$_{5}$N), while a negative correlation with $X$(CH$_{3}$OH).

The $X$(CH$_{3}$CCH) values in the three sources are consistent within their $1 \sigma$ errors, and we cannot find any clear correlations between HC$_{5}$N and CH$_{3}$CCH.
According to the gas-grain-bulk three-phase chemical network simulation \citep{2016MNRAS.458.1859M, 2017MNRAS.467.3525M, 2017MNRAS.466.4470M} results assuming that the temperatures are 20 and 30 K and the density is $10^{5}$ cm$^{-3}$, CH$_{3}$CCH is formed on the grains and desorbed non-thermally \citep{1993MNRAS.261...83H, 2007A&A...467.1103G}.
HC$_{5}$N is formed in the gas phase mainly by the neutral-neutral reaction of CN + C$_{4}$H$_{2}$ and the electron recombination reaction of H$_{2}$C$_{5}$N$^{+}$.
The neutral-neutral reaction is also a main formation pathway to HC$_{5}$N in the model by \citet{2009MNRAS...394...221}.
\citet{2009MNRAS...394...221} suggested that C$_{4}$H$_{2}$ is formed by the reaction of C$_{2}$H$_{2}$ + CCH. 
C$_{2}$H$_{2}$ can be formed in the gas phase from CH$_{4}$ \citep{2008ApJ...681...1385} and can be directly evaporated from grain mantles \citep{2000A&A...355..699L}.
Hence, the productions of CH$_{3}$CCH and HC$_{5}$N seem to be related to the grain-surface reactions even in the lukewarm envelopes ($T \sim 20-30$ K).
No correlation between CH$_{3}$CCH and HC$_{5}$N may imply that there is no direct relationship between them, which is also indicated in the chemical network simulation.

\floattable
\begin{deluxetable}{lcccccc}
\tablecaption{The $N$(H$_{2}$) and fractional abundances of CH$_{3}$OH, CH$_{3}$CCH, HC$_{3}$N, and HC$_{5}$N in the three sources \label{tab:frac}}
\tablewidth{0pt}
\tablehead{
\colhead{Source} & \colhead{$N$(H$_{2}$)\tablenotemark{a}} & \colhead{$X$(CH$_{3}$OH)} & \colhead{$X$(CH$_{3}$CCH)} & \colhead{$X$(HC$_{3}$N)\tablenotemark{b}} & \colhead{$X$(HC$_{3}$N)\tablenotemark{c}} & \colhead{$X$(HC$_{5}$N)\tablenotemark{a}}\\
\colhead{} & \colhead{($\times 10^{22}$ cm$^{-2}$)} & \colhead{($\times 10^{-8}$)} & \colhead{($\times 10^{-8}$)} & \colhead{($\times 10^{-9}$)} & \colhead{($\times 10^{-9}$)} & \colhead{($\times 10^{-10}$)}
}
\startdata
G12.89+0.49 & $2.4^{+2.5}_{-0.8}$ & $12^{+18}_{-9}$ & $4.2^{+2.8}_{-2.2}$ & $1.8 \pm 0.9$ & $5.8^{+4.8}_{-3.6}$ & $9.9^{+6.1}_{-5.4}$ \\
G16.86$-$2.16 & $1.7^{+1.0}_{-0.5}$ & $8.8^{+11}_{-5.8}$ & $3.2^{+1.9}_{-1.5}$ & $2.5^{+1.1}_{-0.9}$ & $7.7^{+5.7}_{-3.9}$ & $16 \pm 7$ \\ 
G28.28$-$0.36 & $0.49^{+0.42}_{-0.16}$ & $4.6^{+7.0}_{-3.4}$ & $7.6^{+5.2}_{-4.4}$ & $4.1^{+2.0}_{-1.9}$ & $16^{+14}_{-9.5}$ & $42^{+26}_{-20}$ \\
\enddata
\tablecomments{The errors represent the standard deviation. These values are averaged ones with the beam size of $18\arcsec$.}
\tablenotetext{a}{\citet{2017ApJ...844...68T}}
\tablenotetext{b}{The HC$_{3}$N column densities were derived from $J=5-4$ and $10-9$.}
\tablenotetext{c}{The HC$_{3}$N column densities were derived from three rotational transitions.}
\end{deluxetable}

\subsection{A Variety of the $N$(HC$_{5}$N)/$N$(CH$_{3}$OH) Ratio}  \label{sec:discuss3}

\begin{figure}
\figurenum{9}
\plotone{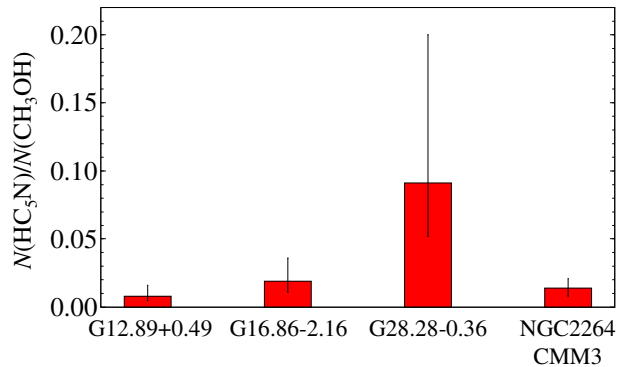}
\caption{Comparison of the $N$(HC$_{5}$N)/$N$(CH$_{3}$OH) ratio among the three high-mass star-forming regions. The data for NGC2264 CMM3 were taken from \citet{2015ApJ...809..162W}. \label{fig:f10}}
\end{figure}

\citet{2017ApJ...844...68T} found that the $N$(HC$_{5}$N)/$W$(CH$_{3}$OH) ratio, where $W$ represents the integrated intensity, varies by more than one order of magnitude among the three sources, and suggested the possibility of the chemical differentiation.
In this paper, we derived the CH$_{3}$OH column densities in the three sources, and then we calculated the $N$(HC$_{5}$N)/$N$(CH$_{3}$OH) ratios in the observed three sources, as shown in Figure \ref{fig:f10}.
We also added the data toward the high-mass protostellar object NGC2264 CMM3 \citep{2015ApJ...809..162W}.
\citet{2015ApJ...809..162W} derived $N$(HC$_{5}$N) and $N$(CH$_{3}$OH) to be ($2.5 \pm 0.9$)$\times 10^{13}$ cm$^{-2}$ and ($1.8 \pm 0.2$)$\times 10^{15}$ cm$^{-2}$, respectively.
We took the cold component value for CH$_{3}$OH, because its rotational temperature ($24.3 \pm 2.6$ K) is comparable with that of HC$_{5}$N ($25.8 \pm 4.6$ K) in NGC2264 CMM3 \citep{2015ApJ...809..162W}.

The $N$(HC$_{5}$N)/$N$(CH$_{3}$OH) ratio in G28.28$-$0.36 ($0.091^{+0.109}_{-0.039}$) is higher than that in G12.89+0.49 ($0.008^{+0.008}_{-0.004}$) by one order of magnitude, and those in G16.86$-$2.16 ($0.019^{+0.017}_{-0.008}$) and NGC2264 CMM3 ($0.014^{+0.007}_{-0.006}$) by a factor of $\sim 5$.

As \citet{2017ApJ...844...68T} discussed, the HC$_{5}$N molecules can be formed from CH$_{4}$ \citep[WCCC,][]{2008ApJ...674...993, 2008ApJ...681...1385} and/or C$_{2}$H$_{2}$ \citep{2009MNRAS...394...221}, which could be evaporated from grain mantles.
CH$_{3}$OH is mainly formed by the successive hydrogenation reaction of CO molecules on dust grains, while CH$_{4}$, may be C$_{2}$H$_{2}$ \citep{2000A&A...355..699L} as well, are formed by the hydrogenation reaction of C atoms on dust grains.
Therefore, the chemical differentiation in the lukewarm envelope may reflect the different ice chemical composition among the MYSOs.

Our source sample was chosen from the HC$_{5}$N-detected source list by \citet{2014MNRAS...443...2252}.
 HC$_{5}$N was not detected in more than half of the sources where \citet{2014MNRAS...443...2252} carried out observations.
 In addition, \citet{2018ApJ...854..133T} detected HC$_{5}$N in half of high-mass protostellar objects in their source list.
 The $N$(HC$_{5}$N)/$N$(CH$_{3}$OH) ratio should cover a broader range of values than the presented here, when considering the HC$_{5}$N-undetected sources into consideration.

\subsection{Relationship between $N$(HC$_{5}$N)/$N$(CH$_{3}$OH) and $N$(CH$_{3}$CCH)/$N$(CH$_{3}$OH)} \label{sec:discuss4}

In this subsection, we follow the analysis by \citet{2015A&A...576A..45F} who compared the chemical composition normalized by CH$_{3}$OH among MYSOs.
We compare the chemical composition in the lukewarm envelope using the $N$(HC$_{5}$N)/$N$(CH$_{3}$OH) and $N$(CH$_{3}$CCH)/$N$(CH$_{3}$OH) ratios.

\begin{figure}
\figurenum{10}
\plotone{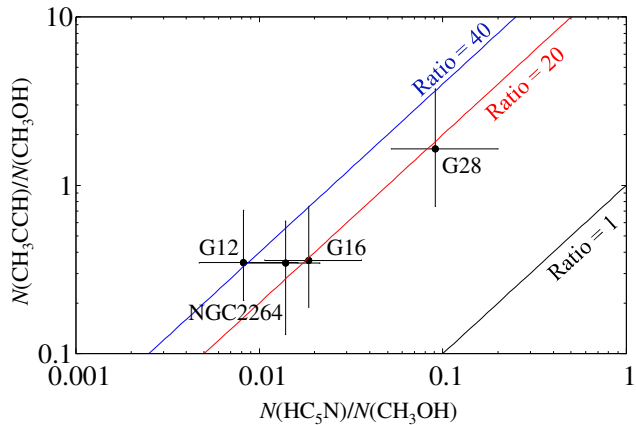}
\caption{Plot of $N$(CH$_{3}$CCH)/$N$(CH$_{3}$OH) versus $N$(HC$_{5}$N)/$N$(CH$_{3}$OH). The data for NGC2264 CMM3 were taken from \citet{2015ApJ...809..162W}. The labels of G12, G16, G28, and NGC2264 indicate the plots for G12.89+0.49, G16.86$-$2.16, G28.28$-$0.36, and NGC2264 CMM3, respectively. \label{fig:f11}}
\end{figure}

Figure \ref{fig:f11} shows the relationship between the $N$(HC$_{5}$N)/$N$(CH$_{3}$OH) ratio and $N$(CH$_{3}$CCH)/$N$(CH$_{3}$OH) ratio in the four MYSOs, the observed three sources and NCG2264 CMM3 \citep{2015ApJ...809..162W}.
The $N$(HC$_{5}$N)/$N$(CH$_{3}$OH) ratio correlates with the $N$(CH$_{3}$CCH)/$N$(CH$_{3}$OH) ratio.
This is caused by the fraction of the lukewarm envelope in the single-dish beams, rather than the chemical differentiation.
The chemical composition, based on HC$_{5}$N and CH$_{3}$CCH, of the lukewarm gas in G28.28$-$0.36 is similar to those in G16.86$-$2.16 and NGC2264 CMM3 (Ratio $\sim 20$).
On the other hand, the $N$(CH$_{3}$CCH)/$N$(CH$_{3}$OH) ratio compared to the $N$(HC$_{5}$N)/$N$(CH$_{3}$OH) ratio in G12.89+0.49 is higher (Ratio $\approx 40$) than those in the other three sources.
These results suggest the chemical diversity in the lukewarm envelope.

\subsection{Relationship of the Chemical Composition between Central Core and Envelope} \label{sec:discuss5}

\citet{2012ApJS..202....1H} carried out 1-mm spectral line survey observations toward 89 GLIMPSE Extended Green Objects \citep[EGOs;][]{2008AJ...136...2391}.
From their survey observations, there are largely two types of EGO clouds; line-rich and line-poor sources.
It is unclear why some sources show line-poor spectra, and others show many lines of COMs.
\citet{2014MNRAS.445.1170G} suggested that the EGO cloud cores are possibly in the short onset phase of the hot core stage, when CH$_{3}$OH ice is quickly evaporating from grain surfaces at a gas temperature of $\sim 100$ K.
Hence, the difference between the line-poor and line-rich EGO cloud cores seems to originate from the chemical differentiation rather than the chemical evolution.
Besides, \citet{2015A&A...576A..45F} compared the chemistry between organic-poor MYSOs and organic-rich MYSOs, namely hot cores, but the relationship between the central core and the envelope was not clear.
In this subsection, we discuss a possibility of the relationship of the chemical composition between the core and the envelope.

There are clear differences in the spectra among the three sources (Figures \ref{fig:f6} $-$ \ref{fig:f5}).
As summarized in Table \ref{tab:resCOMs}, the detected COMs are different among the three observed sources.
CH$_{3}$CHO was detected toward all of the three sources, and it is considered to exist in the envelope \citep{2015A&A...576A..45F}.
In G12.89+0.49, the largest number of COMs have been detected, while G28.28$-$0.36 is a line-poor source.
The results in G28.28$-$0.36, a line-poor source, are consistent with those of \citet{2012ApJS..202....1H}, showing that only H$^{13}$CO$^{+}$ was detected in G28.28$-$0.36.
As discussed in Section \ref{sec:discuss3}, the $N$(HC$_{5}$N)/$N$(CH$_{3}$OH) ratio in G28.28$-$0.36 is higher than that in G12.89+0.49 by an order of magnitude.
In the case of G16.86$-$2.16, the $N$(HC$_{5}$N)/$N$(CH$_{3}$OH) ratio shows a value between G12.89+0.49 and G28.28$-$0.36, and the line density and the COM's line intensities are also between the others.
From these results, the line-poor MYSOs are likely to be surrounded by the carbon-chain-rich envelope, while the line-rich MYSOs, namely hot cores, appear to be surrounded by the CH$_{3}$OH-rich envelope. 

\subsection{Isotopic Fractionation of HC$_{3}$N} \label{sec:discuss6}

The $^{12}$C/$^{13}$C ratio shows a gradient with distance from the Galactic center \citep[$D_{\rm {GC}}$; e.g.,][]{2002ApJ...578...211, 2005ApJ...634...1126}.
Recent observations derived the following relationship \citep{2017ApJ...845..158H}; 
\begin{equation} \label{equ:carbon}
^{12}{\rm {C}}/^{13}{\rm {C}} = 5.21 (\pm 0.52) D_{\rm {GC}} + 22.6 (\pm 3.3),
\end{equation}
where $D_{\rm {GC}}$ is in units of kpc.
We estimated the $D_{\rm {GC}}$ values of each source using trigonometry. 
The $D_{\rm {GC}}$ values are estimated at 5.8 and 8.9 kpc for G12.89+0.49 and G16.86$-$2.16, respectively.
The $D_{\rm {GC}}$ value of G28.28$-$0.36 was derived to be 5.4 kpc by \citet{2016ApJ...830..106T}.
From Equation (\ref{equ:carbon}), the $^{12}$C/$^{13}$C ratios are calculated at $53 \pm 6$, $69 \pm 8$, and $51 \pm 6$ in G12.89+0.49, G16.86$-$2.16, and G28.28$-$0.36, respectively.

The $^{12}$C/$^{13}$C ratios derived from observations (Table \ref{tab:LTE}) are generally lower than or consistent with the values calculated from Equation (\ref{equ:carbon}) within the errors, except for non-detection species.
These results mean that the $^{13}$C species of HC$_{3}$N are not heavily diluted. 
The $^{12}$C/$^{13}$C ratios of cyanopolyynes in dark clouds \citep[e.g., TMC-1, L1521B, and L134N;][]{2016ApJ...817..147T, 2017ApJ...846...46T} are generally lower than those of other carbon-chain species (e.g., CCH, CCS, C$_{3}$S, C$_{4}$H), which means that the $^{13}$C species of cyanopolyynes are not heavily diluted.
Similar results with the $^{12}$C/$^{13}$C ratios of HC$_{3}$N being not significantly high were also found in the warm gas around the protostar L1527 \citep{2016ApJ...830..106T}.
Our results show the similar tendency to the local low-mass star-forming regions.

\citet{2016A&A...587A..91B} reported a tentative detection of DC$_{3}$N toward the Sgr B2 (N2) hot core and derived the D/H ratio to be 0.09\%.
A tentative detection of DC$_{3}$N was also achieved toward the Compact Ridge and the hot core of Orion KL \citep{2013A&A...559A..51E}, and the deuterium fractionation was estimated at $1.5 \pm 0.9$\%.
The deuterium fractionation in the high-mass protostellar object NGC2264 CMM3 was calculated at $1.8 \pm 1.5$\% from a tentative detection of DC$_{3}$N \citep{2015ApJ...809..162W}.
The higher D/H values were reported toward cold dense cores \citep[5\%$-$10\%,][]{1994MNRAS.267...59H} and toward the L1527 protostar, which is one of the warm carbon chain chemistry sources \citep[$\sim 3$\%,][]{2009ApJ...702.1025S}.
The D/H values of HC$_{3}$N in G16.86$-$2.16 and G28.28$-$0.36 ($\sim 1 - 5$\%) is higher than that in Sgr B2, lower than cold dense cores, and comparable to Orion KL hot core and L1527.
On the other hand, DC$_{3}$N was not detected in G12.89+0.49 and we only derived the upper limit for its D/H ratio, which is significantly lower than those in G16.86$-$2.16 and G28.28$-$0.36.
In general, the D/H ratio becomes lower in the higher temperature regions \citep[e.g.,][]{2012A&ARv..20...56C}.
Hence, the results may imply that the emission of HC$_{3}$N comes from the G12.89+0.49 central hot core position, and the lukewarm envelope component is less than those in the other two sources.
In the observations presented here, the lukewarm envelopes and the central cores were covered by the single-dish beam, and we cannot distinguish between them.
The D/H ratio significantly depends on the temperature, and we need the high-spatial resolution observations using interferometers such as ALMA to derive the temperature dependence of the D/H ratio.

\subsection{A Comparison of Line Width of CH$_{3}$OH} \label{sec:discuss1}

Line width is a key tool to characterize the environments where the molecules exist.
In general, high-excitation energy lines are expected to trace the hotter gas and show broad line widths, whereas low-excitation energy lines come mainly from cold molecular clouds and show narrow line widths.
We detected several CH$_{3}$OH emission lines with different excitation energies.
We will investigate the relationship between the excitation energy and the line width of CH$_{3}$OH in this subsection.

\begin{figure}
\figurenum{11}
\plotone{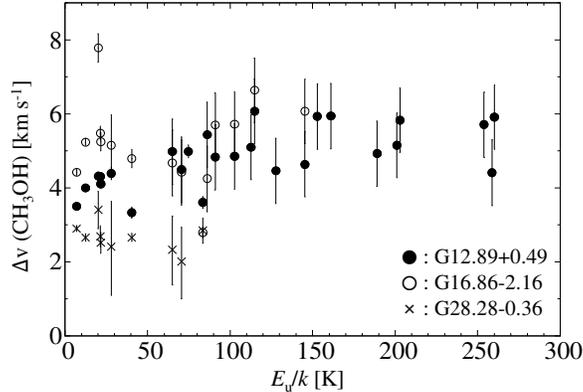}
\caption{Relationship between excitation energy and line width of CH$_{3}$OH. \label{fig:f8}}
\end{figure}

Figure \ref{fig:f8} shows a comparison between the line widths of CH$_{3}$OH and the excitation energy of each line.
We derive the line width ($\Delta v$) using the following formula;
\begin{equation}
\Delta v = \sqrt{\Delta v_{\rm {obs}}^2 - \Delta v_{\rm {inst}}^2},
\end{equation}
where $\Delta v_{\rm {obs}}$ and $\Delta v_{\rm {inst}}$ are the observed line widths (Tables \ref{tab:resCH3OH} and \ref{tab:resASTE}) and the instrumental velocity width (0.85 km s$^{-1}$ $-$ 0.86 km s$^{-1}$, Section \ref{sec:obs}), respectively.

We conducted the Kendall's $\tau$ correlation coefficient test.
The probability ($p$) that there is no correlation between the line width and the excitation energy is calculated at 0.03\%, and the $\tau$ value is $+0.51$ in G12.89+0.49.
This suggests the weak positive correlation between the line width and the excitation energy.
The $p$ and $\tau$ values are derived to be 69\% and $+0.08$ in G16.86$-$2.16, respectively.
In G28.28$-$0.36, the $p$ and $\tau$ values are 72\% and $-0.45$, respectively.
Hence, there is no clear correlation between the line width and the excitation energy in G16.86$-$2.16 and G28.28$-$0.36.
This may be caused by the non-detection of the very-high-excitation energy lines in the two sources.

We also compared the distributions of the CH$_{3}$OH line widths among the three sources using the two-sample Kolmogorov-Smirnov test (K-S test).
We compared the distributions for all of the combinations; (a) G12.89+0.49 and G28.28$-$0.36, (b) G16.86$-$2.16 and G28.28$-$0.36, and (c) G12.89+0.49 and G16.86$-$2.16.
The probabilities that the distributions of the line widths in the selected two sources are the same are derived to be $3.2 \times 10^{-4}$\%, $5.8 \times 10^{-3}$\%, and 39\% for case (a), (b), and (c), respectively.
These results mean that the distribution of the CH$_{3}$OH line width in G28.28$-$0.36 is clearly different from those in the other two sources, and we cannot exclude the possibility that the distributions in G12.89+0.49 and G16.86$-$2.16 are different.

The average line widths of CH$_{3}$OH are $4.8 \pm 0.6$, $5.2 \pm 0.6$, and $2.6 \pm 0.5$ km s$^{-1}$ in G12.89+0.49, G16.86$-$2.16, and G28.28$-$0.36, respectively.
Although the average line widths in G12.89+0.49 and G16.86$-$2.16 are consistent with each other within the errors, G16.86$-$2.16 shows the highest value; nevertheless the very-high-excitation energy lines were not detected in G16.86$-$2.16.
These imply that CH$_{3}$OH exists in turbulent gas such as a molecular outflow in G16.86$-$2.16 rather than the hot gas.
This is supported by the strong wing emission in the CH$_{3}$OH spectra in G16.86$-$2.16.
In G12.89+0.49, CH$_{3}$OH seems to exist in the hot gas as well as in molecular outflow suggested by the wing emission.
On the other hand, the line width in G28.28$-$0.36 is much narrower than those in the other two sources.
This suggests that CH$_{3}$OH exists mainly in the relatively quiescent  region, i.e. envelope for this source.
The non-detection of the very-high-excitation energy lines is consistent with the envelope origin.
There is a possibility of the molecular outflow origin, but a low S/N ratio prevents us from confirmation.

\section{Conclusions}

We carried out observations in the 42$-$46 and 82$-$103 GHz bands with the Nobeyama 45-m radio telescope, and in the 338.2$-$339.2 and 348.45$-$349.45 GHz bands with the ASTE 10-m telescope toward the three high-mass star-forming regions associated with the 6.7 GHz CH$_{3}$OH masers, G12.89+0.49, G16.86$-$2.16, and G28.28$-$0.36.
The rotational temperatures and the beam-averaged column densities of HC$_{3}$N, CH$_{3}$CCH, and CH$_{3}$OH in the three sources are derived.

The $N$(HC$_{5}$N)/$N$(CH$_{3}$OH) ratio in G28.28$-$0.36 is higher than that in G12.89+0.49 by one order of magnitude and that in G16.86$-$2.16 by a factor of 5.
Moreover, the relationships between the $N$(HC$_{5}$N)/$N$(CH$_{3}$OH) ratio and the $N$(CH$_{3}$CCH)/$N$(CH$_{3}$OH) ratio in G28.28$-$0.36 and G16.86$-$2.16 are similar to each other, while HC$_{5}$N is deficient when compared to CH$_{3}$CCH in G12.89+0.49.
These results may imply the chemical diversity of the lukewarm envelope.

The line density in G28.28$-$0.36 is significantly low and a few COMs have been detected, while oxygen-/nitrogen-bearing COMs and high-excitation-energy lines have been detected from G12.89+0.49.
These results seem to imply that the organic-poor MYSOs (G28.28-0.36) are surrounded by the carbon-chain-rich lukewarm envelope, whereas the organic-rich MYSOs (G12.89+0.49 and G16.86-2.16), hot cores, are surrounded by the CH$_{3}$OH-rich lukewarm envelope.
The results presented in this paper were led based on observations with single-dish telescopes without information about the spatial distributions of each molecular emission and only a few molecular species.
Further observations are required to confirm the trends reported in this work.
 



\acknowledgments

We thank the anonymous referee who gave us valuable comments which helped us improve the quality of this paper.
We would like to express our great thanks to the members of the Nobeyama Radio Observatory.
The Nobeyama Radio Observatory is a branch of the National Astronomical Observatory of Japan (NAOJ), National Institutes of Natural Sciences.
The Z45 receiver is supported in part by a Granting-Aid for Science Research of Japan (24244017).
We thank to the operation staff members of the ASTE.
The ASTE telescope is operated by the NAOJ.
KT appreciates support from a Granting-Aid for Science Research of Japan (17J03516).
LM also acknowledges support from the NASA postdoctoral program.  
A portion of this research was carried out at the Jet Propulsion Laboratory, California Institute of Technology, under a contract with the National Aeronautics and Space Administration.



\vspace{5mm}
\facilities{Nobeyama 45-m radio telescope, Atacama Submillimeter Telescope Experiment (ASTE)}
\software{Java NEWSTAR (https://www.nro.nao.ac.jp/~nro45mrt/html/obs/newstar/index.html)}


\begin{thebibliography}{}
\bibitem[Aikawa et al.(2008)]{2008ApJ...674...993} Aikawa, Y., Wakelman, V., Garrod, R. T., \& Herbst, E.\ 2008, \apj, 674, 993
\bibitem[Bacmann et al.(2012)]{2012A&A...541L..12B} Bacmann, A., Taquet, V., Faure, A., Kahane, C., \& Ceccarelli, C.\ 2012, \aap, 541, L12
\bibitem[Balucani et al.(2015)]{2015MNRAS.449L..16B} Balucani, N., Ceccarelli, C., \& Taquet, V.\ 2015, \mnras, 449, L16
\bibitem[Belloche et al.(2016)]{2016A&A...587A..91B} Belloche, A., M{\"u}ller, H.~S.~P., Garrod, R.~T., \& Menten, K.~M.\ 2016, \aap, 587, A91 
\bibitem[Bisschop et al.(2007)]{2007A&A...465..913B} Bisschop, S.~E., J{\o}rgensen, J.~K., van Dishoeck, E.~F., \& de Wachter, E.~B.~M.\ 2007, \aap, 465, 913
\bibitem[Caselli \& Ceccarelli(2012)]{2012A&ARv..20...56C} Caselli, P., \& Ceccarelli, C.\ 2012, \aapr, 20, 56
\bibitem[Cernicharo et al.(2012)]{2012ApJ...759L..43C} Cernicharo, J., Marcelino, N., Roueff, E., et al.\ 2012, \apjl, 759, L43
\bibitem[Chapman et al.(2009)]{2009MNRAS...394...221} Chapman, J. F., Millar, T. J., Wardle, M., Burton, M. G., \& Walsh, A. J.\ 2009, \mnras, 394, 221
\bibitem[Cyganowski et al.(2008)]{2008AJ...136...2391} Cyganowski, C. J., Whitney, B. A., Holden, E., et al.\  2008, \aj, 136, 2391
\bibitem[Esplugues et al.(2013)]{2013A&A...559A..51E} Esplugues, G.~B., Cernicharo, J., Viti, S., et al.\ 2013, \aap, 559, A51
\bibitem[Fayolle et al.(2015)]{2015A&A...576A..45F} Fayolle, E.~C., {\"O}berg, K.~I., Garrod, R.~T., van Dishoeck, E.~F., \& Bisschop, S.~E.\ 2015, \aap, 576, A45 
\bibitem[Garrod et al.(2007)]{2007A&A...467.1103G} Garrod, R.~T., Wakelam, V., \& Herbst, E.\ 2007, \aap, 467, 1103
\bibitem[Ge et al.(2014)]{2014MNRAS.445.1170G} Ge, J.~X., He, J.~H., Chen, X., \& Takahashi, S.\ 2014, \mnras, 445, 1170 
\bibitem[Goldsmith \& Langer(1999)]{1999ApJ...517..209G} Goldsmith, P.~F., \& Langer, W.~D.\ 1999, \apj, 517, 209
\bibitem[Green et al.(2014)]{2014MNRAS...443...2252} Green, C. -E., Green, J. A., Burton, M. G., et al.\  2014, \mnras, 443, 2252
\bibitem[Halfen et al.(2017)]{2017ApJ...845..158H} Halfen, D.~T., Woolf, N.~J., \& Ziurys, L.~M.\ 2017, \apj, 845, 158
\bibitem[Hasegawa \& Herbst(1993)]{1993MNRAS.261...83H} Hasegawa, T.~I., \& Herbst, E.\ 1993, \mnras, 261, 83
\bibitem[Hassel et al.(2008)]{2008ApJ...681...1385} Hassel, G. E., Herbst, E., \& Garrod, R. T.\ 2008, \apj, 681, 1385
\bibitem[He et al.(2012)]{2012ApJS..202....1H} He, J.~H., Takahashi, S., \& Chen, X.\ 2012, \apjs, 202, 1
\bibitem[Herbst \& van Dishoeck(2009)]{2009ARA&A..47..427H} Herbst, E., \& van Dishoeck, E.~F.\ 2009, \araa, 47, 427 
\bibitem[Hirota et al.(2009)]{2009apj...699...585} Hirota, T., M. Ohishi, \& Yamamoto, S\  2009, \apj, 699, 585
\bibitem[Howe et al.(1994)]{1994MNRAS.267...59H} Howe, D.~A., Millar, T.~J., Schilke, P., \& Walmsley, C.~M.\ 1994, \mnras, 267, 59
\bibitem[Iguchi \& Okuda(2008)]{2008PASJ...60..857I} Iguchi, S., \& Okuda, T.\ 2008, \pasj, 60, 857 
\bibitem[Imai et al.(2016)]{2016ApJ...830L..37I} Imai, M., Sakai, N., Oya, Y., et al.\ 2016, \apjl, 830, L37
\bibitem[Jaber et al.(2014)]{2014ApJ...791...29J} Jaber, A.~A., Ceccarelli, C., Kahane, C., \& Caux, E.\ 2014, \apj, 791, 29
\bibitem[Kamazaki et al.(2012)]{2012PASJ...64...29K} Kamazaki, T., Okumura, S.~K., Chikada, Y., et al.\ 2012, \pasj, 64, 29
\bibitem[Lahuis \& van Dishoeck(2000)]{2000A&A...355..699L} Lahuis, F., \& van Dishoeck, E.~F.\ 2000, \aap, 355, 699
\bibitem[Li et al.(2016)]{2016aj...152...92L} Li, F. C., Xu, Y., Wu, Y. W., et al.\  2016, \aj, 152, 92
\bibitem[Majumdar et al.(2017a)]{2017MNRAS.467.3525M} Majumdar, L., Gratier, P., Andron, I., Wakelam, V., \& Caux, E.\ 2017a, \mnras, 467, 3525
\bibitem[Majumdar et al.(2017b)]{2017MNRAS.466.4470M} Majumdar, L., Gratier, P., Ruaud, M., et al.\ 2017b, \mnras, 466, 4470
\bibitem[Majumdar et al.(2016)]{2016MNRAS.458.1859M} Majumdar, L., Gratier, P., Vidal, T., et al.\ 2016, \mnras, 458, 1859
\bibitem[Marcelino et al.(2007)]{2007ApJ...665L.127M} Marcelino, N., Cernicharo, J., Ag{\'u}ndez, M., et al.\ 2007, \apjl, 665, L127
\bibitem[Milam et al.(2005)]{2005ApJ...634...1126} Milam, S. N., Savage, C., Brewster, M. A., Ziurys, L. M., \& Wyckoff, S.  2005, \apj, 634, 1126
\bibitem[M$\ddot{\rm u}$ller et al.(2005)]{2005JMoSt...742...215} M$\ddot{\rm u}$ller, H. S. P., Schl$\ddot{\rm o}$der, F., Stutzki, J., \& Winnewisser, G.\  2005, JMoSt, 742, 215
\bibitem[Nakajima et al.(2013)]{2013PASP..125..252N} Nakajima, T., Kimura, K., Nishimura, A., et al.\ 2013, \pasp, 125, 252
\bibitem[Nakamura et al.(2015)]{2015PASJ...67..117N} Nakamura, F., Ogawa, H., Yonekura, Y., et al.\ 2015, \pasj, 67, 117
\bibitem[{\"O}berg et al.(2010)]{2010ApJ...716..825O} {\"O}berg, K.~I., Bottinelli, S., J{\o}rgensen, J.~K., \& van Dishoeck, E.~F.\ 2010, \apj, 716, 825
\bibitem[Okuda \& Iguchi(2008)]{2008PASJ...60..315O} Okuda, T., \& Iguchi, S.\ 2008, \pasj, 60, 315
\bibitem[Potapov et al.(2016)]{2016A&A...594A.117P} Potapov, A., S{\'a}nchez-Monge, {\'A}., Schilke, P., et al.\ 2016, \aap, 594, A117 
\bibitem[Purcell et al.(2006)]{2006MNRAS...367...553} Purcell, C. R., Balasubramanyam, R., Burton, M. G., et al.\  2006, \mnras, 367, 553
\bibitem[Purcell et al.(2009)]{2009MNRAS.394..323P} Purcell, C.~R., Longmore, S.~N., Burton, M.~G., et al.\ 2009, \mnras, 394, 323
\bibitem[Reboussin et al.(2014)]{2014MNRAS.440.3557R} Reboussin, L., Wakelam, V., Guilloteau, S., \& Hersant, F.\ 2014, \mnras, 440, 3557
\bibitem[Reid et al.(2014)]{2014apj...783...130} Reid, M. J., Menten, K. M., Brunthaler, A., et al.\  2014, \apj, 783, 130
\bibitem[Ruaud et al.(2016)]{2016MNRAS.459.3756R} Ruaud, M., Wakelam, V., \& Hersant, F.\ 2016, \mnras, 459, 3756
\bibitem[Sakai et al.(2009a)]{2009ApJ...697..769S} Sakai, N., Sakai, T., Hirota, T., Burton, M., \& Yamamoto, S.\ 2009a, \apj, 697, 769
\bibitem[Sakai et al.(2008)]{2008ApJ...672..371S} Sakai, N., Sakai, T., Hirota, T., \& Yamamoto, S.\ 2008, \apj, 672, 371-381
\bibitem[Sakai et al.(2009b)]{2009ApJ...702.1025S} Sakai, N., Sakai, T., Hirota, T., \& Yamamoto, S.\ 2009b, \apj, 702, 1025 
\bibitem[Savage et al.(2002)]{2002ApJ...578...211} Savage, C., Apponi, A. J., Ziurys, L. M., \& Wyckoff, S.  2002, \apj, 578, 211
\bibitem[Shimajiri et al.(2015)]{2015ApJS..221...31S} Shimajiri, Y., Sakai, T., Kitamura, Y., et al.\ 2015, \apjs, 221, 31
\bibitem[Suzuki et al.(1992)]{1992apj...392...551} Suzuki, H., Yamamoto, S., Ohishi, M., et al.\  1992, \apj, 392, 551
\bibitem[Taniguchi et al.(2017a)]{2017ApJ...846...46T} Taniguchi, K., Ozeki, H., \& Saito, M.\ 2017a, \apj, 846, 46
\bibitem[Taniguchi et al.(2016a)]{2016ApJ...817..147T} Taniguchi, K., Ozeki, H., Saito, M., et al.\ 2016a, \apj, 817, 147
\bibitem[Taniguchi et al.(2017b)]{2017ApJ...844...68T} Taniguchi, K., Saito, M., Hirota, T., et al.\ 2017b, \apj, 844, 68
\bibitem[Taniguchi et al.(2016b)]{2016ApJ...830..106T} Taniguchi, K., Saito, M., \& Ozeki, H.\ 2016b, \apj, 830, 106
\bibitem[Taniguchi et al.(2018)]{2018ApJ...854..133T} Taniguchi, K., Saito, M., Sridharan, T.~K., \& Minamidani, T.\ 2018, \apj, 854, 133
\bibitem[Tielens \& Hagen(1982)]{1982A&A...114..245T} Tielens, A.~G.~G.~M., \& Hagen, W.\ 1982, \aap, 114, 245
\bibitem[Urquhart et al.(2013)]{2013MNRAS.431.1752U} Urquhart, J.~S., Moore, T.~J.~T., Schuller, F., et al.\ 2013, \mnras, 431, 1752
\bibitem[van Dishoeck(2017)]{2017arXiv171005940V} van Dishoeck, E.~F.\ 2017, arXiv:1710.05940
\bibitem[Vastel et al.(2014)]{2014ApJ...795L...2V} Vastel, C., Ceccarelli, C., Lefloch, B., \& Bachiller, R.\ 2014, \apjl, 795, L2
\bibitem[Watanabe et al.(2015)]{2015ApJ...809..162W} Watanabe, Y., Sakai, N., L{\'o}pez-Sepulcre, A., et al.\ 2015, \apj, 809, 162
\end{thebibliography}
\end{document}